\documentclass{aa}
\usepackage{graphicx}

\newcommand{\ratioo} {N({\rm H}_2) / I_{\rm CO}}
\def\la{\lower.5ex\hbox{$\; \buildrel < \over \sim \;$}}
\def\ga{\lower.5ex\hbox{$\; \buildrel > \over \sim \;$}}

\begin{document}      

   \title{A dynamical model for the Taffy galaxies UGC 12914/5}

   \author{B.~Vollmer\inst{1}, J.~Braine\inst{2,3}, \& M.~Soida\inst{4}}

   \offprints{B.~Vollmer, e-mail: Bernd.Vollmer@astro.unistra.fr}

   \institute{CDS, Observatoire astronomique, UMR 7550, 11, rue de l'universit\'e,
	      67000 Strasbourg, France \and
	      Univ. Bordeaux, Laboratoire d'Astrophysique de Bordeaux, UMR 5804, F-33270, Floirac, France \and
	      CNRS, LAB, UMR 5804, F-33270, Floirac, France \and
	      Astronomical Observatory, Jagiellonian University, ul. Orla 171, PL-30-244 Krak\'ow, Poland
              }

   \date{Received / Accepted}

   \authorrunning{Vollmer, Braine, \& Soida}
   \titlerunning{A Taffy model}

\abstract{
The spectacular head-on collision of the two gas-rich galaxies of the Taffy system, UGC~12914/15, gives us
a unique opportunity to study the consequences of a direct ISM-ISM collision.
To interpret existing multi-wavelength observations, we made dynamical simulations of the Taffy system including a sticky particle component.
To compare simulation snapshots to H{\sc i} and CO observations, we assume that the molecular fraction of the gas depends on the
square root of the gas volume density. For the comparison of our simulations with observations
of polarized radio continuum emission, we calculated the evolution of the 3D large-scale magnetic field for our simulations.
The induction equations including the time-dependent gas-velocity fields from the dynamical model were solved for this purpose.
Our simulations reproduce the stellar distribution of the primary galaxy, UGC~12914, the prominent H{\sc i} and CO gas bridge,
the offset between the CO and H{\sc i} emission in the bridge, the bridge isovelocity vectors parallel to the bridge,
the H{\sc i} double-line profiles in the bridge region, the large line-widths ($\sim 200$~km\,s$^{-1}$) in the bridge region,
the high field strength of the bridge large-scale regular magnetic field, the projected magnetic field vectors parallel to 
the bridge and the strong total power radio continuum emission from the bridge. 
The stellar distribution of the secondary model galaxy is more perturbed than observed.
The observed distortion of the H{\sc i} envelope of the Taffy system is not reproduced by our simulations which use initially symmetric gas disks.
The model allows us to define the bridge region in three dimensions. We estimate the total bridge gas mass (H{\sc i}, warm and cold H$_{2}$)
to be $5$ to $6 \times 10^{9}$~M$_{\odot}$, with a molecular fraction $M_{\rm H_{2}}/M_{\rm HI}$ of about unity.
Despite the enormous mass of molecular gas in the bridge, very little star formation is present,
similar to other systems with extraplanar gas and broad CO lines.
The structure of the model gas bridge is bimodal: on kpc-scales there is a dense ($\ga 0.01$~M$_{\odot}$pc$^{-3}$) component
with a high velocity dispersion $> 100$~km\,s$^{-1}$ and a less dense ($\sim 10^{-3}$~M$_{\odot}$pc$^{-3}$)
component with a smaller, but still high, velocity dispersion $\sim 50$~km\,s$^{-1}$. The synchrotron lifetime of
relativistic electrons is only long enough to be consistent with the existence of the radio continuum bridge 
for the less dense component. 
On the other hand, only the high-density gas undergoes a high enough mechanical energy input to produce the observed strong emission
of warm H$_{2}$. We propose that, despite the high local gas densities, this high input of mechanical energy drives strong turbulence and 
quenches star formation in the bridge gas except for the giant H{\sc ii} region near UGC~12915.
Our model suggests that we observe this galaxy head-on collision near the time of maximum CO and H$_{2}$ emission.
}

\maketitle

\section{Introduction \label{sec:intro}}

Galactic evolution is largely due to interactions of various sorts.
We can distinguish tidal encounters, which are essentially gravitational interactions,
from head-on collisions in which the gaseous components actually hit each other, 
resulting in fantastic shocks and injection of energy into the gas.  A further type of
interaction is the ram-pressure stripping suffered by cluster galaxies.

We are concerned here with the head-on collisions and focus on the Taffy Galaxies,
UGC~12914 and UGC~12915, a spectacular bridge system first noticed by Condon et al. (1993). 
In these collisions, the stellar morphology is
determined by the tidal interaction while the morphology of the gaseous component
is determined by both gravity and hydrodynamics.  While the general geometry of the system was
suggested by Condon et al. (1993), no numerical simulations of the collision have been made until now.

The Taffy system attracted attention through its strong radio synchrotron bridge, a very 
unusual feature.  The bridge is HI-rich and was subsequently found to be rich in molecular 
gas as well through CO observations (Gao et al. 2003, Braine et al. 2003).  Dust appears to be underabundant
with respect to gas in the bridge (Zink et al. 2000, Zhu et al. 2007), presumably due to grain ablation
during the collision.  The galaxies themselves are particularly massive, with rotation velocities
of $250$~km\,s$^{-1}$ or more and some $1.5 \times 10^{10}$~M$_{\odot}$ of H{\sc i} and a similar quantity of 
molecular gas, dependent on the $\ratioo$ conversion factor from CO emission to H$_2$ column
density.  Some 10 -- 20\,\% of the gas is in the bridge, making it at least as rich in gas as the 
entire Milky Way. Condon et al. (1993) estimate that the galaxy disks passed through each other about 20~Myr
ago with a transverse velocity of about $600$~km\,s$^{-1}$.  The bridge and the counter-rotating galaxies 
are seen close to edge-on as the recession velocities of the two galaxies are virtually equal. Fig.~\ref{fig:img20} 
shows the morphology of the system.

The wealth of available data makes the Taffy system both attractive and challenging to model.
We present the first simulations of the collision, attempting to simultaneously reproduce 
($i$) the morphologies of the stellar and gaseous components, 
($ii$) the relative bridge and galaxy gas masses, 
($iii$) the velocity field including the presence of double peaks in the bridge spectra,
($iv$) the magnetic field orientation, and 
($v$) the morphology of the polarized and unpolarized radio continuum emission.
A sticky particle code is used in order to enable cloud-cloud collisions to occur.

We are most familiar with the interstellar medium (ISM) and star formation (SF) in 
the Milky Way, a quiescent rotating disk. At the velocities observed in the Milky Way, 
ISM-ISM (i.e. hydro) collisions are believed to favor
star formation, the clearest example being the triggered star formation sometimes found around 
H{\sc ii} regions (Zavagno et al. 2010).  
The impact velocity of the interstellar media of the UGC~12914/5 system is $\ga 800$~km\,s$^{-1}$, 
very much greater than can be observed in the Milky Way. Despite the 
enormous gas mass and strong CO emission in the bridge, strikingly little star formation is 
present (Braine et al. 2004).  Numerically modeling the encounter is the first step towards understanding
the effect of injecting huge amounts of kinetic energy into the gas found in the bridge.

The UGC~12914/5 system is not the only ISM-ISM collision known.   The so-called Taffy2 system, 
UGC~813/6 (Condon et al. 2002), is very similar although the galaxies are somewhat less massive and 
less gas-rich.  Nonetheless, the bridge is quite spectacular and the geometry quite similar.
Stephan's Quintet is similar in terms of collision velocity but the velocity is entirely along the line of sight;
the collision is on-going and involves an intruder galaxy hitting the intra-group medium of this
compact group with a velocity of $\sim 1000$~km\,s$^{-1}$. The resulting large-scale shock ($\sim 40$~kpc) 
emits in X-ray (Trinchieri et al. 2003) and radio continuum emission (van der Hulst 1981).
Powerful high-velocity dispersion molecular hydrogen is associated with the intergalactic shock wave
(Appleton et al. 2006, Guillard et al. 2009). About $5 \times 10^{8}$~M$_{\odot}$ of warm 
H$_{2}$ spread over $\sim 480$~kpc$^{2}$ were found in the main shock region (Cluver et al. 2010).
In addition, CO(1--0), (2--1) and (3--2) line emission have been detected in this region with complex profiles, 
spanning a velocity range $\sim 1000$~km\,s$^{-1}$ (Guillard et al. 2012).
The intra-group material involved has most probably been tidally stripped by a past galaxy--galaxy interaction
(Renaud et al. 2010, Hwang et al. 2012). Colliding ring galaxies can also have gas and star bridges.
As an example, the H{\sc i} bridge between the two galaxies in Arp~284 (NGC~7714/15) contains $2 \times 10^{9}$~M$_{\odot}$ 
(Smith et al. 1997). Struck \& Smith (2003) showed that a fast ($\sim 400$~km\,s$^{-1}$) off-center ($\sim 4$~kpc) 
inclined collision is responsible for the peculiar morphology of the system. The models suggest that the gas bridge interacts
with tidally stripped gas from an older component.

Head-on collisions of gas-rich galaxies are rare events in the local Universe. However,
galaxy encounters of all types were more frequent, and galaxies more gas-rich, at earlier epochs.
The Taffy system UGC~12914/15 can thus give insight into how mass assembled in the early Universe.

\section{The model \label{sec:model}}

We used the N-body code described in Vollmer et al. (2001), which consists of two components: a non-collisional component that 
simulates the stellar bulge/disk and the dark halo, and a collisional component that simulates the ISM.
The non-collisional component of each galaxy consists of 81œôòé920 particles, which simulate the galactic halo, bulge, and disk. 
The characteristics of the different galactic components are presented in Table~\ref{tab:parameters}.
The first of the two galaxies, the primary, is $2.3$ times more massive and its disk scalelength is $1.6$ times larger than that of
the secondary galaxy.
The resulting rotation velocities are $\sim 295$~km\,s$^{-1}$ and $\sim 245$~km\,s$^{-1}$, respectively.

We adopted a model where the ISM is simulated as a collisional component, i.e. as discrete particles that possess a mass and a 
radius and can have partially inelastic collisions. Since the ISM is a turbulent and fractal medium (see e.g. Elmegreen \& Falgarone 1996), 
it is neither continuous nor discrete. The volume filling factor of the warm and cold phases is smaller than one. The warm neutral 
and ionized gas fills about $30$-$50$\,\% of the volume, whereas cold neutral gas has a volume filling factor smaller than $10$\,\% 
(Boulares \& Cox 1990). It is unclear how this fraction changes when an external pressure is applied. 
We thus do not identify the cloud particles with giant molecular clouds (GMC) but treat the entire ISM as a collisional medium.
In contrast to smoothed 
particle hydrodynamics (SPH), which is a quasi-continuous approach where the particles cannot penetrate each other, our approach 
allows a finite penetration length, which is given by the mass-radius relation of the particles. 

The 20\,000 particles of the collisional component in each galaxy represent gas cloud complexes that evolve in the gravitational 
potential of the galaxy. The total assumed gas mass of the two galaxies is $2.4 \times 10^{10}$~M$_{\odot}$.
The mass distribution of the cloud complexes is $N(m) {\rm d}m \propto m^{-1.3} {\rm d}m $ for 
$2 \times 10^{5}$~M$_{\odot} \leq m \leq 2 \times 10^{7}$~M$_{\odot}$. This is close to the mass spectrum of Galactic
giant molecular clouds ($m^{-1.5}$; Solomon et al. 1987, Rosolowsky 2005). To each particle, a radius is attributed depending on its mass:
\begin{equation}
\label{eq:xi}
r= 65\ \sqrt{\xi (\frac{m}{10^{6}~{\rm M}_{\odot}})}\ {\rm pc.}
\end{equation}
We used different values for the cloud size parameter $\xi$, which is not known a priori. 

\begin{figure}
  \centering
  \resizebox{\hsize}{!}{\includegraphics{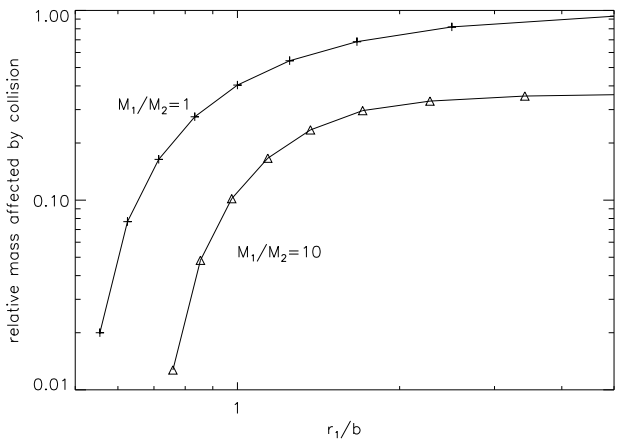}}
  \caption{Overlapping cloud mass which is affected by a cloud-cloud collision divided by the total cloud mass as a function 
    of the fraction between the radius of the massive cloud $r_{1}$ and the impact parameter $b$ (from Wiegel 1994). 
    Crosses: cloud of equal mass $M_{1}=M_{2}$. Triangles: $M_{1}= 10 \times M_{2}$.
  \label{fig:collisions}}
\end{figure}

During the disk evolution, the cloud particles can have partially 
inelastic collisions, the outcome of which (coalescence, mass exchange, or fragmentation) is simplified following
the geometrical prescriptions of Wiegel (1994). Fig.~\ref{fig:collisions} shows the overlapping mass fraction for a collision
of two clouds as a function of the ratio between the cloud radius $r_{\rm cl}$ and the impact parameter $b$.
For a collision of clouds of equal mass, the ratio between the overlapping and the total cloud masses
is small for $b \sim 2 r_{\rm cl}$ and rises steeply to $\sim 0.5$ for $b=r_{\rm cl}$. For $b < r_{\rm cl}$
the mass fraction slowly rises toward unity. For a collision of clouds with a mass ratio of ten,
the shape of the overlapping mass fraction with respect to $r_{1}/b$ is the same as in the previous case, but
shifted to higher $r_{1}/b$ and lower mass fractions. We conclude that for $r_{1}< b < r_{1}+r_{2}$ the mass fraction
which is involved in the cloud-cloud collision is small.
Our method which treats the cloud collisions geometrically is close to that of Olson \& Kwan (1990) and Theis \& Hensler (1993). 
All partially inelastic collisions conserve mass and momentum. The collisions result in an effective gas viscosity in the disk.
Since the code has a maximum number of particles of 40\,000, mass exchange between the two colliding gas clouds occurs most frequently.
Colliding clouds can only fragment if cloud coalescence occurred before elsewhere. During a mass exchange the less massive cloud
always gains and the more massive cloud loses mass. After a large number of collisions, this leads to steepening of the mass
spectrum at the high and low-mass ends, and ultimately to a flat mass spectrum around the mean cloud mass of $6 \times 10^{5}$~M$_{\odot}$.

The particle trajectories are integrated using an adaptive timestep for each particle. This method is described in 
Springel et al. (2001). The following criterion for an individual timestep is applied:
\begin{equation}
\Delta t_{i}= \frac{20\ {\rm km\,s}^{-1}}{a_{i}}
\end{equation}
where $a_{i}$ is the acceleration of the particle $i$. The minimum value of $t_{i}$ defines the global timestep used for the 
Burlisch-Stoer integrator that integrates the collisional component, and is typically a few $10^{4}$~yr. 

Since each cloud-cloud collision is resolved in time and space by the code, the cloud collision rate is physical and
follows the classical equation
\begin{equation}
\label{eq:collrate}
\frac{{\rm d}N}{{\rm d}t} = \frac{2}{\sqrt{\pi}} \sum_{i=1}^{N}\ v_{{\rm disp},i} \sigma_{i} n_{i},
\end{equation}
where $v_{{\rm disp},i}$ is the 1D velocity dispersion, $\sigma_{i}$ the cross section of particle $i$, and $n_{i}$ is the local cloud density 
around particle $i$ (Theis \& Hensler 1993). 
Since we use a cloud mass spectrum, the cross section is $\sigma = \epsilon \pi r^{2}$ , with $1 \leq \epsilon \leq 4$, 
instead of $\sigma= 4\,\pi r^{2}$ that applies for particles of same size. 
For the comparison with the collision rate of the dynamical simulations in Sect.~\ref{sec:simulations}, we set $\epsilon = 2$.

\begin{table}
      \caption{Pre-collision model galaxy parameters.}
         \label{tab:parameters}
      \[
         \begin{array}{lccc}
           \hline
           \noalign{\smallskip}
           {\rm component} & {\rm mass} & N^{(1)} & r_{\rm d}^{(2)} \\
	    & (10^{10}~{\rm M}_{\odot}) & & {\rm (kpc)}  \\
	   \noalign{\smallskip}
	   \hline
	   \noalign{\smallskip}
	   {\rm galaxy\ 1} & v_{\rm rot}=295 & {\rm km\,s}^{-1} & \\
	   \noalign{\smallskip}
	   \hline
	   \noalign{\smallskip}
	   {\rm halo} & 64.3 & 32768 & 10.0  \\
	   {\rm bulge} & 2.2  & 16384 & 0.9  \\
	   {\rm stellar\ disk} & 11.1 & 32768 & 4.0  \\
	   {\rm gas\ disk} & 1.7 & 20000 & 6.0  \\
	   \noalign{\smallskip}
	   \hline
	   \noalign{\smallskip}
	   {\rm galaxy\ 2} & v_{\rm rot}=245 & {\rm km\,s}^{-1} & \\
	   \noalign{\smallskip}
	   \hline
	   \noalign{\smallskip}
	   {\rm halo} & 28.2 & 32768 & 6.3  \\
	   {\rm bulge} & 1.0  & 16384 & 0.6  \\
	   {\rm stellar\ disk} & 4.9 & 32768 & 2.5  \\
	   {\rm gas\ disk} & 0.7 & 20000 & 3.8  \\
	   \noalign{\smallskip}
	   \hline
        \end{array}
      \]
      \begin{list}{}{}
      \item[$^{(1)}$Number of particles;]
      \item[$^{(2)}$exponential scale length.]
      \end{list}
\end{table}

\section{The simulations \label{sec:simulations}}

We made two sets of simulations of head-on collisions of two gas-rich spiral galaxies:
(i) a first set with a simplified collision geometry to determine the cloud size parameter $\xi$ and the inclination angle $i$
between the two galaxies before the collision and (ii) a second set of simulations with fixed $\xi$ and
a fixed range of inclinations, where we varied the impact parameter systematically. 

\subsection{Cloud size parameter and galaxy inclination \label{sec:first}}

In the first set of simulations the more massive (primary) galaxy is placed in the $x$-$y$ plane, the second galaxy
is placed in a plane parallel to that of the first one, but offset by $12$~kpc along the $z$-axis
and by $0.5$~kpc along the $x$-axis. The initial velocity of the less massive galaxy is $500$~km\,s$^{-1}$
along the $z$-axis toward the primary galaxy which is at rest.
We varied the inclination angle between the two galactic disks between $-45^{\circ}$ and $45^{\circ}$ and
the cloud size parameter between $1 \leq \xi \leq 16$ (Table~\ref{tab:simulations}).

For the sky projection we used the observed position and inclination angles of UGC~12914: $PA=160^{\circ}$, $i=30^{\circ}$. 
The position angle and inclination of the model primary galaxy define a plane in three dimensional space.
The model galaxy can then be rotated within this plane by the azimuthal viewing angle which is chosen such
that the secondary galaxy is close to the observed position of UGC~12915.
\begin{figure}
  \centering
  \resizebox{\hsize}{!}{\includegraphics{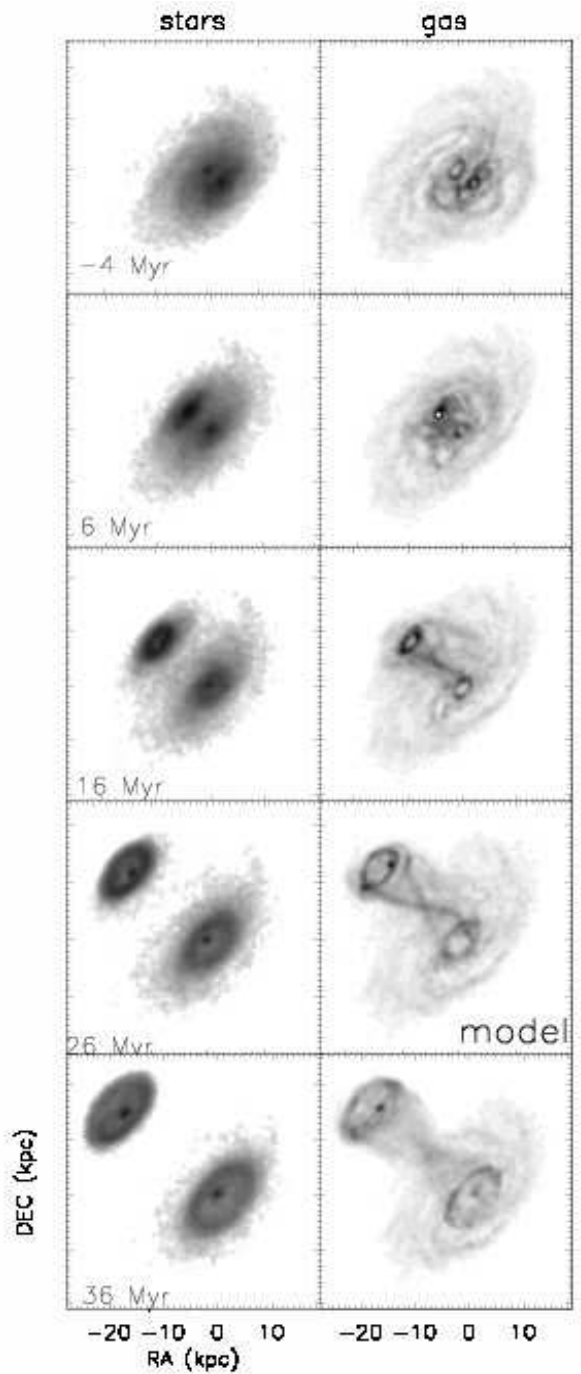}}
  \caption{Time evolution of a Taffy galaxy collision (simulation 10). Left panels: stellar surface density.
    The greyscale is logarithmic from 8 to 5600~M$_{\odot}$pc$^{-2}$.
    Right panels: total gas surface density. The greyscale are $(1,4,9,16,25,36,49,...,400)$~M$_{\odot}$pc$^{-2}$.
    The timestep of interest (26~Myr) is marked by ``model''. The impact parameter is $\sim 3$~kpc.
  \label{fig:TAFFY12_evol}}
\end{figure}
Fig.~\ref{fig:TAFFY12_evol} shows a representative time evolution of the simulated head-on collision between two 
gas-rich galaxies. The galaxies collide\footnote{We define the collision time as the time of minimum distance between the galaxy centers.} 
at $t=0$~Myr. The projected distance between the two galaxies is
close to the distance between UGC~12914 and UGC~12915 at $t=26$~Myr. The head-on collision produces a prominent
gas bridge between the galaxies without a stellar counterpart. At $t=26$~Myr the gas within the bridge forms an X-structure.
The outer gas disk of the primary galaxy is not affected by the collision because of the small size of the initial
gas disk of the secondary galaxy. 
For $t \geq 20$~Myr both galaxies develop stellar and gaseous ring structures which are characteristic for head-on galaxy 
collisions (see, e.g., the Cartwheel galaxy; Higdon et al. 1995, 1996).
\begin{figure*}
  \centering
  \resizebox{16cm}{!}{\includegraphics{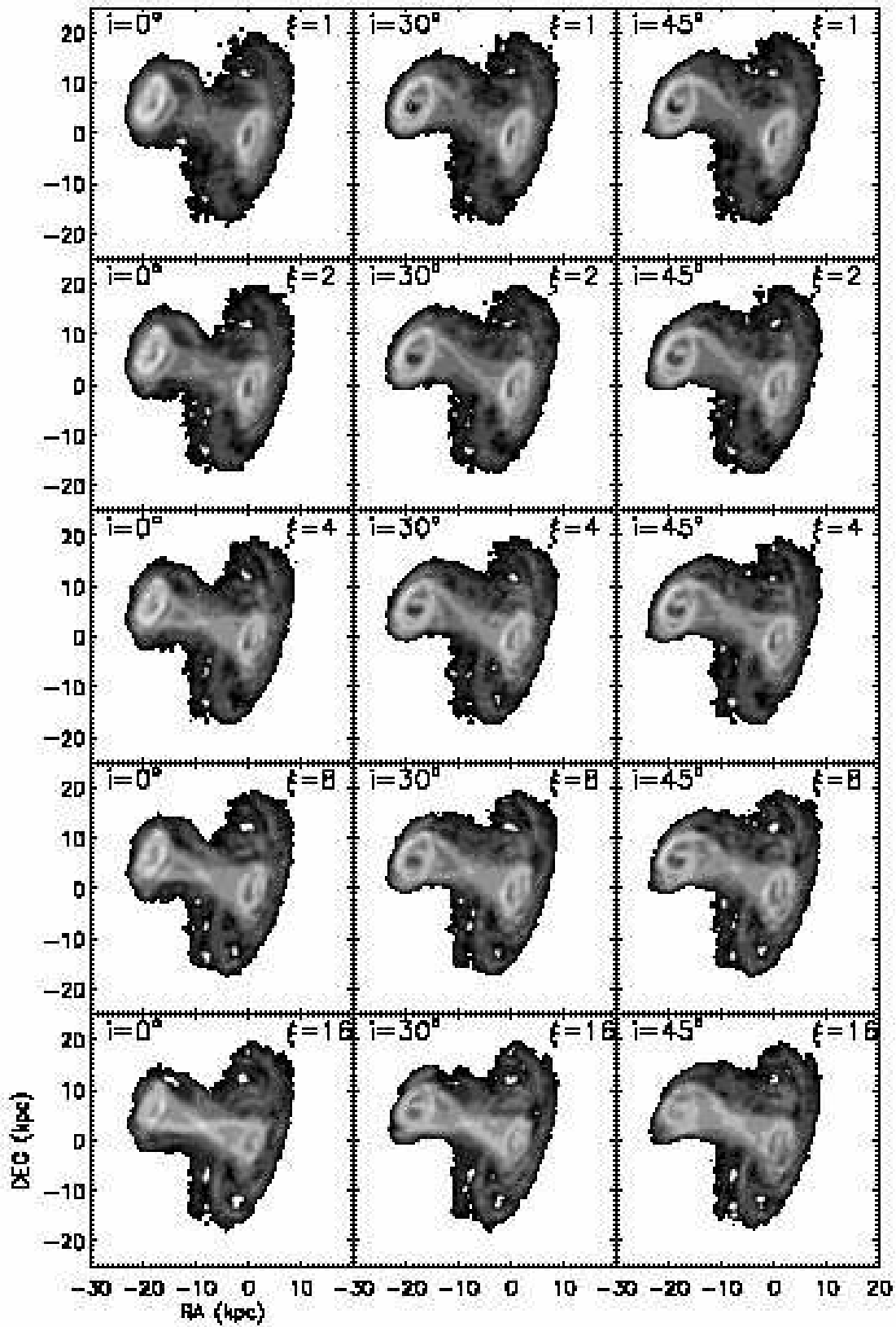}}
  \caption{Snapshots at the time of interest (26~Myr) for simulations with different inclination angles between
    the disks and different cloud particle sizes as described in Table~\ref{tab:simulations}.
    Greyscale levels are $(2,4,8,16,32,64,128,256)$~M$_{\odot}$pc$^{-2}$.
  \label{fig:simulations}}
\end{figure*}

\begin{table}
      \caption{Parameters of the first set of Taffy simulations.}
         \label{tab:simulations}
      \[
         \begin{array}{lrrrr}
           \hline
           \noalign{\smallskip}
           {\rm Taffy} & \ \ \ \ \ \ \xi^{(1)} & {\rm inclination}^{(2)}  & N_{\rm coll}^{\rm max\ (3)} & M_{\rm bridge}^{(4)} \\
	   {\rm simulation} & &  & ({\rm Myr}^{-1}) & (10^{9}~{\rm M}_{\odot}) \\
	   \noalign{\smallskip}
	   \hline
	   \noalign{\smallskip}
	   1 & 1 & 0^{\circ} &  891 &    1.48 \\
	   2 & 1 & 30^{\circ} & 893 &     1.50 \\
	   3 & 1 & 45^{\circ} & 880 &     1.98 \\
	   4 & 2 & 0^{\circ} &  1525 &     1.79 \\
	   5 & 2 & 30^{\circ} & 1609 &     1.87 \\
	   6 & 2 & 45^{\circ} & 1579 &     2.29 \\
	   7 & 4 & 0^{\circ} & 2609  &    2.41 \\
	   8 & 4 & 30^{\circ} &  2362 &     2.12 \\
	   9 & 4 & 45^{\circ} & 2280  &    2.48 \\
	   10 & 8 & 0^{\circ} & 3249  &    2.75 \\
	   11 & 8 & 30^{\circ} & 3218  &    2.64 \\
	   12 & 8 & 45^{\circ} & 2945  &    2.77 \\
	   13 & 8 & -30^{\circ} &  3793 &     3.19 \\
	   14 & 8 & -45^{\circ} & 4600 &      3.33 \\
	   15 & 16 & 0^{\circ} & 5623  &    4.01 \\
	   16 & 16 & 30^{\circ} & 5046 &     3.48 \\
	   17 & 16 & 45^{\circ} & 4963 &     3.62 \\
	   \noalign{\smallskip}
	   \hline
        \end{array}
      \]
      \begin{list}{}{}
      \item[$^{(1)}$Parameter of the cloud mass-size relation;]
      \item[$^{(2)}$inclination angle between the two disks;]
      \item[$^{(3)}$maximum collision rate between cloud particles;]
      \item[$^{(4)}$total gas mass within the bridge region $3 < z < 12$~kpc] 
      \end{list}
\end{table}

The gas distribution of the bridge displays an X-structure in all simulations.
Since the X-structure is visible at all azimuthal viewing angles, we conclude that it
stems from a 3D hourglass shape. 
The X-structure is symmetric for $i=0^{\circ}$, i.e. when the colliding disks are parallel.
Increasing the inclination angle between the two disks increases the gas surface density of the
upper part of the X-structure, toward the less massive galaxy leading to a higher bridge mass.
An increase of the cloud size parameter $\xi$ leads to an increase of the overall 
gas surface density in the bridge region without changing the gas morphology.

\begin{figure}
  \centering
  \resizebox{\hsize}{!}{\includegraphics{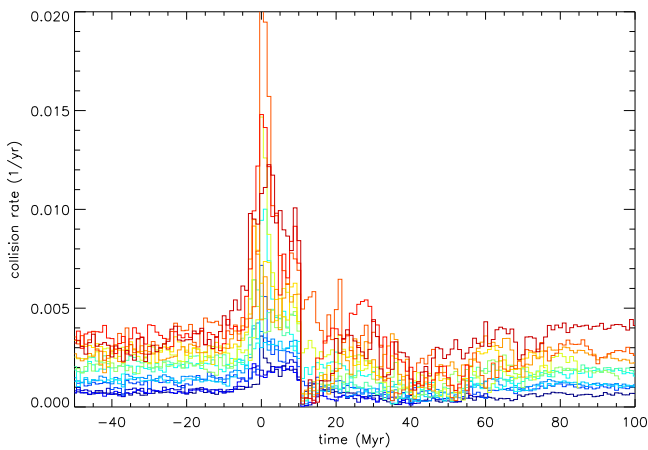}}
  \caption{Cloud particle collision rate for the simulations described in Table~\ref{tab:simulations}.
    The factor of the cloud particle mass-size relation $\xi$ increases from blue to red.
  \label{fig:collisions_graph}}
\end{figure}

The time evolution of the cloud collisions in the whole system are presented in Fig.~\ref{fig:collisions_graph}.
The collision rate of the clouds in the two galactic disks is constant before the collision.
As expected, its absolute value increases with increasing cloud size parameter $\xi$ (from blue to red).
During the collision, the cloud collision rate rises steeply within a few Myr, stays high during
$\sim 10$~Myr, drops to a minimum at $t \sim 40$~Myr, and rises again slightly for $t > 40$~Myr (Fig.~\ref{fig:collisions_graph}).
The latter rise is due to the formation of prominent gas rings of high surface densities in both galaxies.
The ratio between the maximum and initial collision rates mainly depends on the inclination angle
between the disks. This fraction is $\sim 13$ for $i=0^{\circ}$ and $\sim 22$ for $i=30^{\circ}$-$45^{\circ}$.

\begin{figure}
  \centering
  \resizebox{\hsize}{!}{\includegraphics{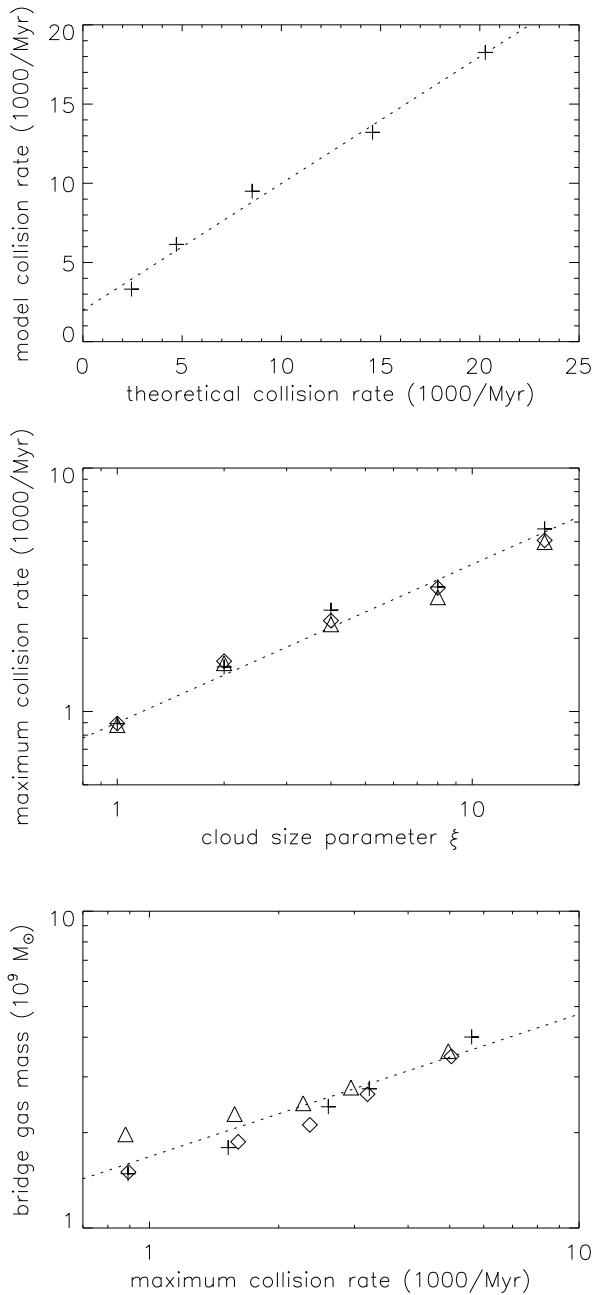}}
  \caption{Correlation for the simulations described in Table~\ref{tab:simulations}.
    Upper panel: cloud collision rate of the simulated quiet galaxies before collision as a function of the
    theoretical cloud collision rate based on Eq.~\ref{eq:collrate}.
    Middle panel: maximum collision rate in bins of $10$~Myr as a function of the cloud particle mass-size relation $\xi$.
    Lower panel: total gas mass in the bridge region as a function of the maximum collision rate.
    Crosses: $i=0^{\circ}$; diamonds: $i=30^{\circ}$; triangles: $i=45^{\circ}$.
  \label{fig:statistics}}
\end{figure}

The quantitative relations of the influence of the inclination angle between the disks $i$ and
the cloud size parameter $\xi$ on the cloud collision rate during the galaxy collision and the
gas mass in the bridge region are shown in Fig.~\ref{fig:statistics}. We define the bridge region as
a cylinder whose base is parallel to the disk plane of the primary galaxy with an infinite radius and extending from $3$~kpc to $12$~kpc
perpendicular to the plane of the primary galaxy (UGC~12914). 
Based on Eq.~\ref{eq:collrate}, the collision rate of an ensemble of particles  
should be proportional to the cloud size parameter $\xi$ (Eq.~\ref{eq:xi}). The situation changes if the ensemble has an 
evolving spectrum of particle sizes. After a large number of collisions, the mass spectrum steepens at the high end
(see Sect.~\ref{sec:model}). This steepening decreases the number of 
massive clouds which dominate the overall collision rate because of their large cross sections.
The local particle density and velocity dispersion is calculated using the $50$ nearest neighbors.
As expected for a slight steepening of the cloud mass spectrum with increasing collision rate, 
the correlation between the theoretical collision rate of the system based on Eq.~\ref{eq:collrate} and the
measured model collision rate of the quiet galaxy system before collision is close to linear with a slope of $0.8$
(upper panel of Fig.~\ref{fig:statistics}). 
The ratio between the maximum collision rate counted in bins of $10$~Myr and the collision rate of the quiet system  
is $5.3 \pm 0.6$ for all simulations. The correlation between the cloud size parameter $\xi$ and the maximum collision rate
has a slope of $0.65$ (middle panel of Fig.~\ref{fig:statistics}), i.e. it is somewhat shallower than the correlation of the collision 
rate of the quiescent system. We ascribe this difference again to the cloud size spectrum.

The correlation between the bridge gas mass and the maximum collision rate is 
$M_{\rm bridge} \propto ({\rm d}N/{\rm d}t)_{\rm max}^{0.5}$ (lower panel of Fig.~\ref{fig:statistics}).
This behavior is caused by the geometrical cloud collision scheme (Sect.~\ref{sec:model}): 
for a constant impact parameter $b$, the ratio between the mass of the colliding clouds that is affected by the
collision $m_{\rm coll}$, i.e. the mass included in the geometrical overlap between the clouds, and the initial masses of the colliding 
clouds $m_{\rm ini}$ is small for $r_{1} < b < r_{1}+r{2}$ and increases rapidly with increasing cloud radius $r_{1}$.
Only cloud-cloud collisions with $b/r_{1} < 0.7$ lead to significant transfer of mass and momentum.
An increase of the cloud radius with increasing $\xi$ (Eq.~\ref{eq:xi}) thus leads to an increase of the collision rate which is
approximately proportional to $\xi$. However, many of the new cloud-cloud collisions do not lead to a significant mass and
momentum transfer. Since the gas bridge is caused by momentum transfer during cloud-cloud collisions, the bridge gas mass
is expected to increase more slowly than the cloud collision rate, as observed in our simulations.

With $({\rm d}N/{\rm d}t)_{\rm max} \propto \xi^{0.65}$, the correlation between the bridge gas mass and the cloud size parameter $\xi$ is 
$M_{\rm bridge} \propto \xi^{0.3}$. The maximum collision rate does not depend on the inclination angle
between the disks. The dependence of the bridge mass on the inclination angle between the disk
is weak compared to that on the cloud size parameter $\xi$ (Fig.~\ref{fig:statistics}).

For $\xi \geq 8$ the bridge gas mass exceeds $2.7 \times 10^{9}$~M$_{\odot}$ (Table~\ref{tab:simulations}). This is at the lower end, 
but comparable to the observed bridge gas mass of $3$-$4 \times 10^{9}$~M$_{\odot}$. We thus decided to fix the cloud size parameter
at $\xi = 8$ for the second set of simulations. Since only simulations with $-30^{\circ} \leq i \leq 0^{\circ}$
lead to distributions of stars and gas that reproduce observations, we restricted the second set of simulations
to this range of inclination angles between the two disks.

\subsection{Comparison with SPH simulations}

The main difference between our collisional model and SPH simulations of head-on collisions (Struck 1997, Tsuchiya et al. 1998)
is the finite penetration length of the collisional particles compared to the interdiction of penetration in SPH.
The impact generally carves a hole in the primary disk which is filled quite rapidly (Struck 1997).
Since in SPH all particles in the overlapping region collide, one expects a much higher momentum transfer in SPH than in a collisional scheme.
Due to the strong momentum transfer, a less massive ($\sim 20$\% of the primary) companion gas disk is destroyed during the 
head-on collision (Struck 1997).
As in our model, head-on collisions of gas-rich galaxies form prominent gas bridges in SPH.
The SPH bridges also show high gas column densities, comparable to those found in our simulations.
In contrast with our model, SPH simulations can include gas heating and cooling. Struck (1997) showed that the colliding gas
is heated to temperatures up to $10^{6}$~K at the impact. The bridge gas than cools down to $\sim 10^{5}$~K after $35$~Myr and  
to $\sim 10^{4}$~K within less than $100$~Myr\footnote{A large fraction of the Taffy bridge gas is 
cooler. Cooling strongly depends on gas density. SPH models have kpc resolution
and do not resolve dense small-scale structures, which cool on much shorter timescales.}. 
The overall gas distributions of isothermal SPH simulations compared to
SPH simulation including heating and cooling are similar. In general, the edges of the gas distributions are sharper
in the isothermal simulations (Fig.~4 of Struck 1997). We expect that our model simulations are closer to
isothermal SPH simulations of head-on galaxy collisions, but they involve a smaller amount of momentum transfer between the ISM
of the colliding galaxies.

\subsection{Realistic Taffy galaxy models \label{sec:second}}

\begin{figure*}
  \centering
  \resizebox{8cm}{!}{\includegraphics{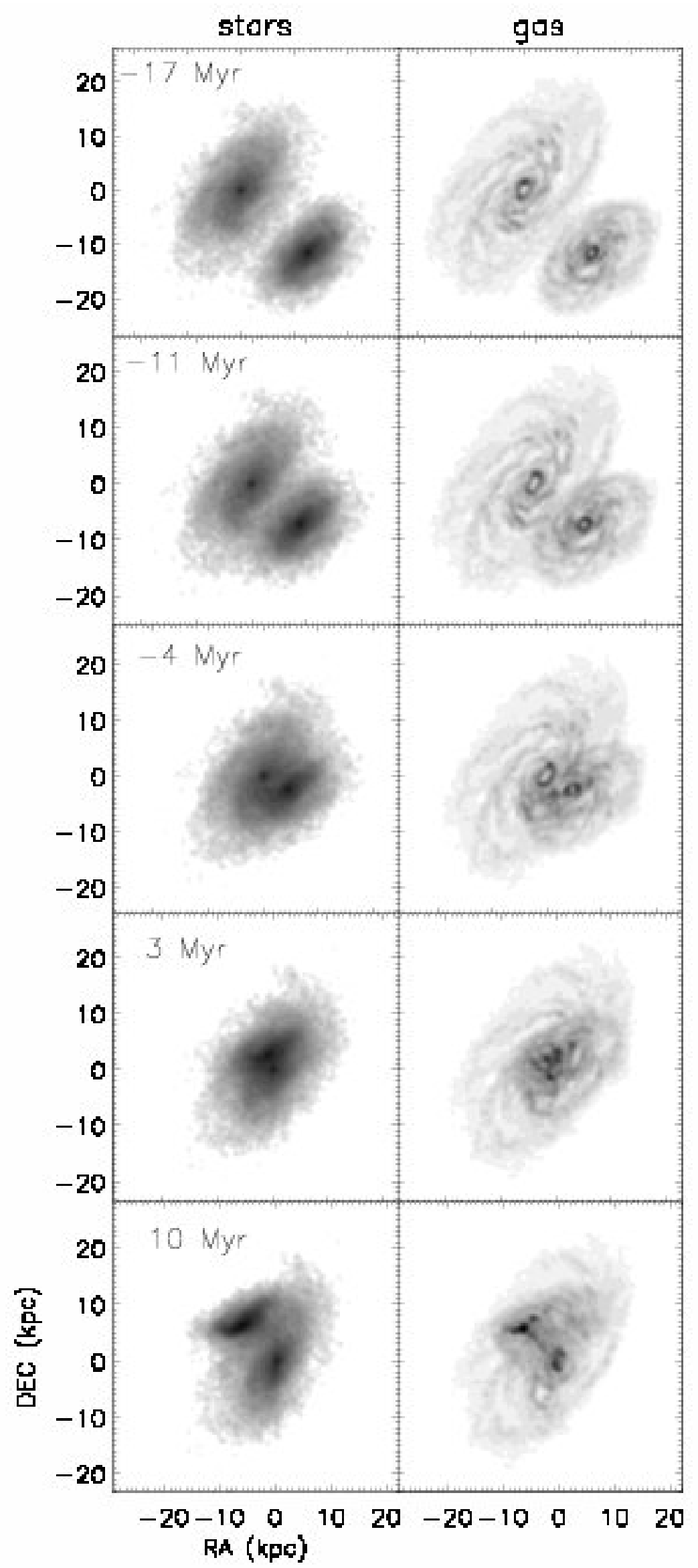}}
  \resizebox{8cm}{!}{\includegraphics{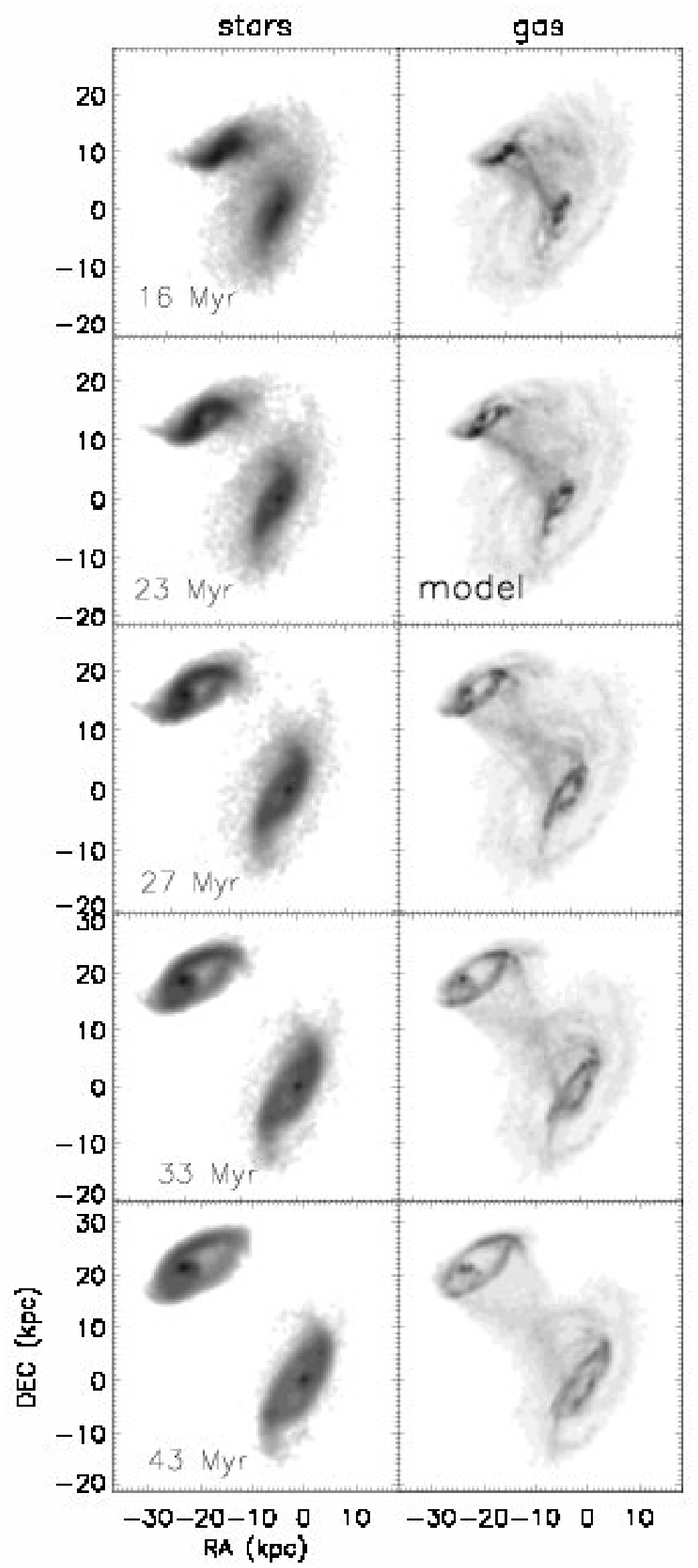}}
  \caption{Time evolution of simulation 20 ($i=0^{\circ}$, $\delta y = 4$~kpc). Left panels: stellar surface density.
    The greyscale is logarithmic from 8 to 5600~M$_{\odot}$pc$^{-2}$.
    Right panels: total gas surface density. The greyscale are $(1,4,9,16,25,36,49,...,400)$~M$_{\odot}$pc$^{-2}$.
    The timestep of interest (26~Myr) is marked by ``model''.
  \label{fig:TAFFY26new2_evol}}
\end{figure*}

\begin{figure*}
  \centering
  \resizebox{8cm}{!}{\includegraphics{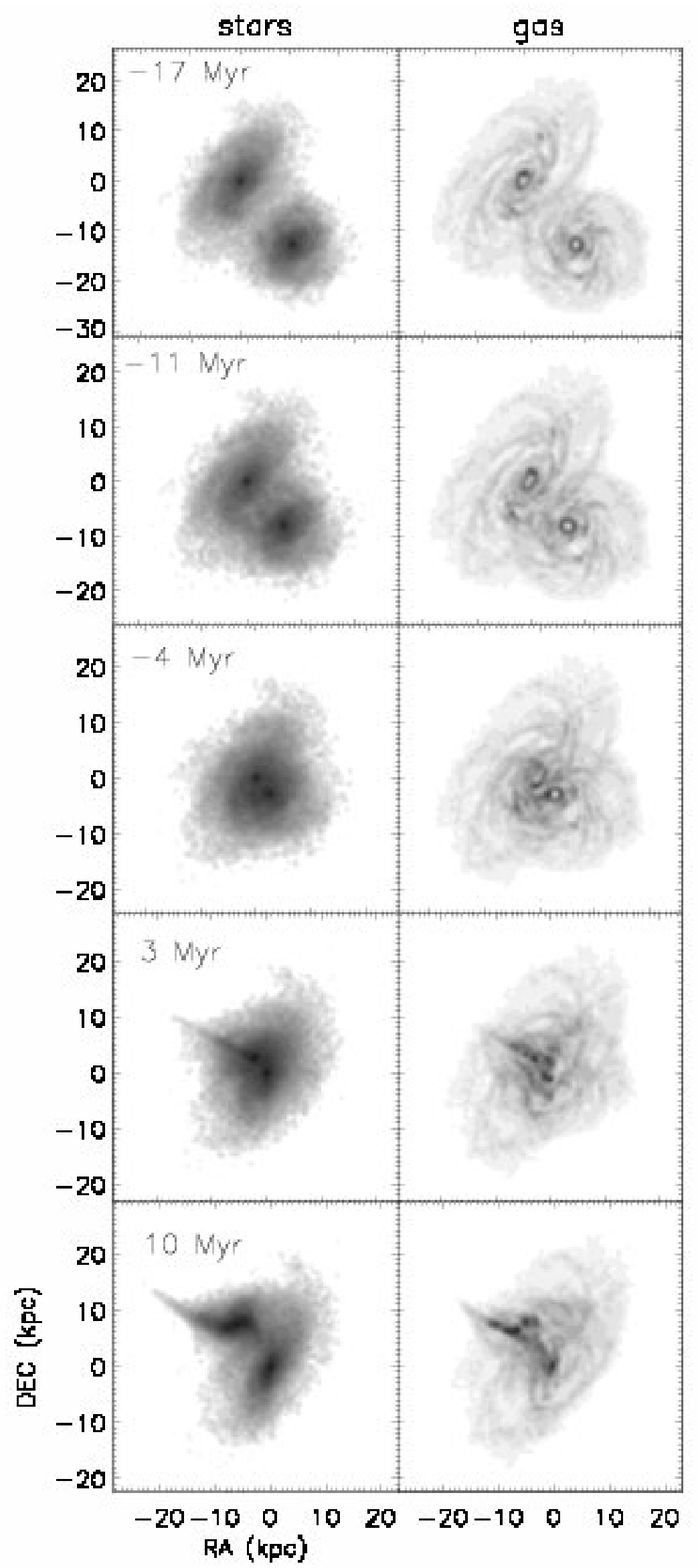}}
  \resizebox{8cm}{!}{\includegraphics{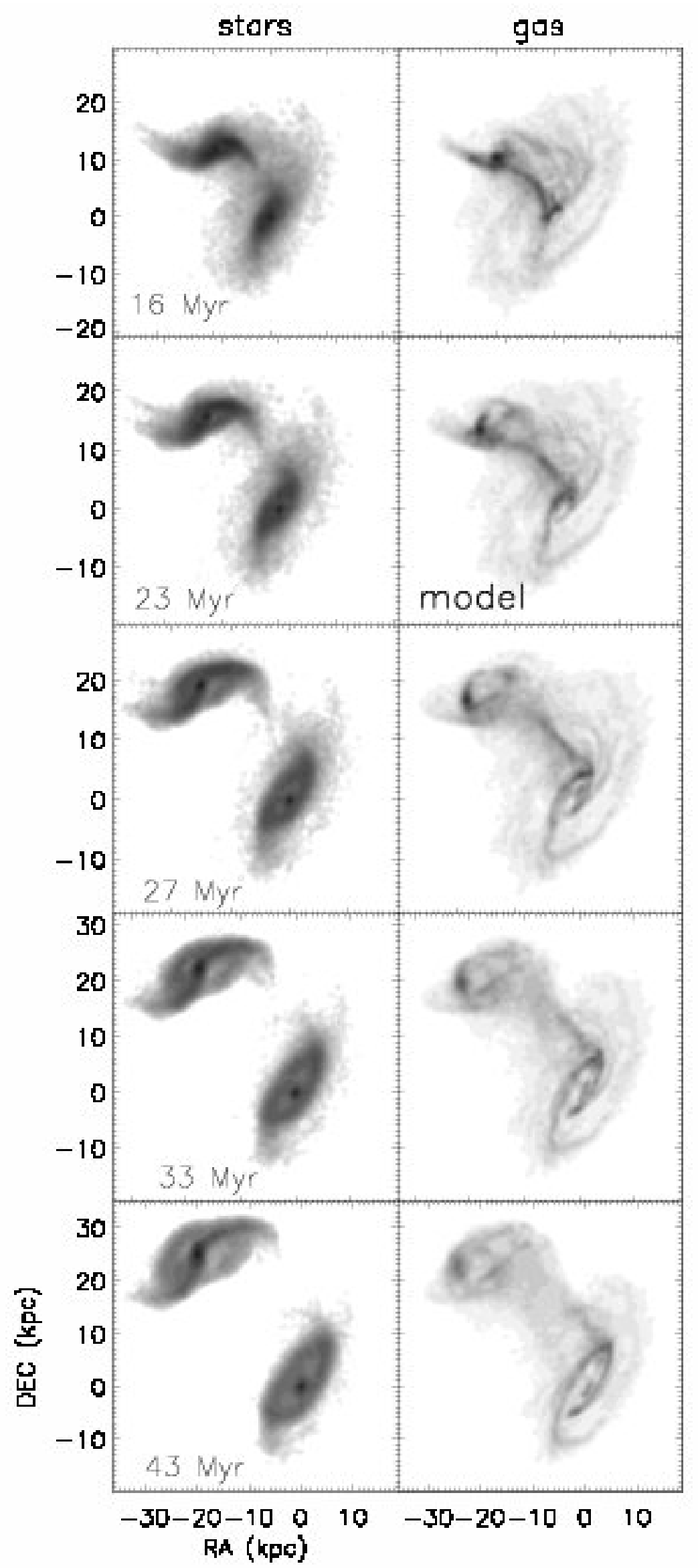}}
  \caption{Time evolution of simulation 19 ($i=-30^{\circ}$, $\delta y = 1$~kpc). Left panels: stellar surface density.
    The greyscale is logarithmic from 8 to 5600~M$_{\odot}$pc$^{-2}$.
    Right panels: total gas surface density. The greyscale are $(1,4,9,16,25,36,49,...,400)$~M$_{\odot}$pc$^{-2}$.
    The timestep of interest (26~Myr) is marked by ``model''.
  \label{fig:TAFFY22new_evol}}
\end{figure*}

In our second set of simulations we set the cloud size parameter $\xi=8$ and the inclination between the two galaxies 
$i=-30^{\circ},\ 0^{\circ}$. To test the influence on the impact parameter, we varied the initial position of the
secondary galaxy $(-71.5,1.0,-96.5)$~kpc by adding $3$~kpc-long and $9$~kpc-long vectors with directions pointing toward the
eight corners and six sides of a cube. From these $28$ simulations we chose those which best fit the observed
CO gas distribution and velocity field. The initial conditions of these four simulations are presented in Table~\ref{tab:simulations1}.
\begin{table*}
      \caption{Parameters of the second set of Taffy simulations which are used for comparison with observations.}
         \label{tab:simulations1}
      \[
         \begin{array}{lcccccccccccc}
           \hline
           \noalign{\smallskip}
           {\rm Taffy} & \ \ \ \ \ \ \xi & {\rm inclination}  & x & y & z & v_{x} & v_{y} & v_{z} & b & \Delta v_{\rm max} & N_{\rm coll}^{\rm max} & M_{\rm bridge} \\
	   {\rm simulation} & & & {\rm (kpc)} & {\rm (kpc)} & {\rm (kpc)} & {\rm (km\,s}^{-1}) & {\rm (km\,s}^{-1}) & {\rm (km\,s}^{-1}) & ({\rm kpc}) & {\rm (km\,s}^{-1}) & ({\rm Myr}^{-1}) & (10^{9}~{\rm M}_{\odot}) \\
	   \noalign{\smallskip}
	   \hline
	   \noalign{\smallskip}
	   18 & 8 & 0^{\circ} & -71.5 & 1.0 & -96.5 & 300 & 0 & 405 & 2.0 & 1080 & 4850 & 3.49 \\
	   19 & 8 & -30^{\circ} &-71.5 &1.0 &-96.5 & 300 & 0 & 405 & 1.0 & 1120 & 4950 & 3.22 \\ 
	   20 & 8 & 0^{\circ} &-71.5 & 4.0 &-96.5 & 300 & 0 & 405 & 2.5 & 1080 & 3548 & 3.21 \\
	   21 & 8 & -30^{\circ} &-71.5 &4.0 &-96.5  & 300 & 0 & 405 & 1.3 & 1120 & 4973 & 3.23 \\
	   \noalign{\smallskip}
	   \hline
        \end{array}
      \]
\end{table*}
In the following, we restrict our discussion on the two ``best-fit'' models, simulations~19 and 20.
These simulations differ in the inclination angle between the disks, the impact parameter, and the maximum relative velocity.
Whereas the maximum relative velocity between the two galaxies is between $1080$ and $1120$~km\,s$^{-1}$, it decreases rapidly after
the impact, and is only between $660$ and $690$~km\,s$^{-1}$ at the time of interest, i.e. $23$~Myr after the impact.
The time evolution of the stellar and gaseous components for are presented in Figs.~\ref{fig:TAFFY26new2_evol} (simulation 20) and 
\ref{fig:TAFFY22new_evol} (simulation 19). 
The main difference between these simulations is that in simulation~19 the secondary galaxy forms strong tidal tails to the
east and west parallel to the secondary's trajectory, much stronger than observed. The origin of a part of the gas mass in the bridge region is the
secondary's western tidal arm. These tails are also formed in simulation~20, but they are much shorter 
($3$~kpc instead of $10$~kpc for simulation~19).

Both simulations produce prominent gas bridges along the secondary's trajectory whose morphologies are very similar to
those of the previous set of simulations with simplified collision geometries (Fig.~\ref{fig:TAFFY12_evol}).
Whereas the gas bridge of simulation~20 has the morphology of a 3D hourglass, the morphology of the gas bridge of simulation~19
is dominated by a straight filament which bifurcates toward the secondary galaxy. For both simulations the high-surface density
small-scale structures disperse and disappear $\sim 30$~Myr after the galaxy collision. At the same time, the width of the gas
bridge increases. As in the previous set of
simulations, the outer gas disk of the primary galaxy is not affected by the interaction, because of the relatively small
extent of the secondary's gas disk. For $t > 30$~Myr, both galaxies develop prominent ring stellar and gaseous structures.

\begin{figure}
  \centering
  \resizebox{\hsize}{!}{\includegraphics{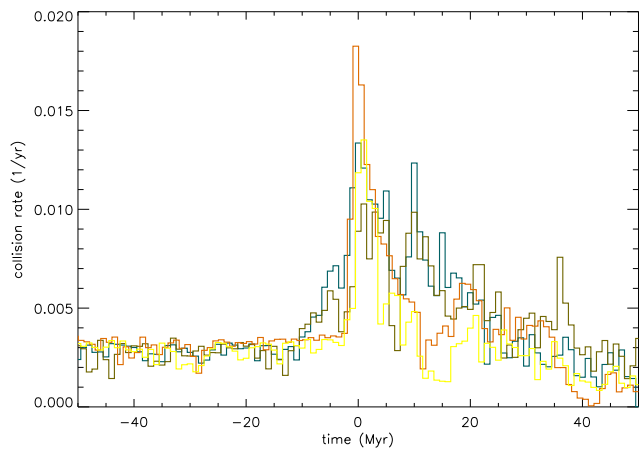}}
  \caption{Cloud particle collision rate for the simulations described in Table~\ref{tab:simulations1}.
    Red: simulation 18; blue: simulation 19; yellow: simulation 20; green: simulation 21.
  \label{fig:collisions_graph1}}
\end{figure}

The total cloud collision rates of the system for simulations~18--21 are shown in Fig.~\ref{fig:collisions_graph1}.
The shape of the time evolution of the cloud collision rate mainly depends on the inclination angle $i$ between
the two disks. In the case of parallel disks ($i=0^{\circ}$, simulations~18 and 20), the cloud collision rate
abruptly increases during a few Myr. After its peak, it decreases until $10$--$15$~Myr,
rises again slightly until $20$--$25$~Myr, and then declines slowly until $50$~Myr after the impact.
The collision rate shows a different behavior for inclined galactic disks ( $i=-30^{\circ}$, simulations~19 and 21):
the cloud collision rate slowly rises already $10$~Myr before the impact. At the moment of impact, it rises
abruptly during a few Myr. After the impact, the collision rate declines until $t \sim 7$~Myr, increases again to
a secondary maximum at $t \sim 10$~Myr, and then declines monotonically until $t=50$~Myr.
The formation of the secondary maximum is due to the formation of the prominent tidal tail in the secondary galaxy
(see timestep $10$~Myr in Fig.~\ref{fig:TAFFY22new_evol}). For $t > 20$~Myr the evolution of the collision rates
are similar for all four simulations.

\subsection{The time of impact}

The time of interest is chosen to reproduce the main observed characteristics of the Taffy system:
the projected positions of UGC~12914/15, the high gas column density of the bridge, and the double-line profile
of the spectra of the total gas in the bridge region. This leads us to choose the timestep $23$~Myr after impact. 
This timestep of interest is indicated in Figs.~\ref{fig:TAFFY26new2_evol} and \ref{fig:TAFFY22new_evol}

Condon et al. (1993) estimated the time of impact under the assumption that their orbits are parabolic. 
Since the systemic velocities of UGC~12914 and UGC~12915 are very close ($\Delta v_{\rm r} = 35$~km\,s$^{-1}$),
the galaxies 3D velocity vectors are located in the plane of the sky. 
With the projected distance between the galaxies ($\sim 15$~kpc) and galaxy masses, they derived a relative
velocity of $600$~km\,s$^{-1}$. This leads to a time since impact of $t \sim 24$~Myr which is consistent with
the time derived from spectral steepening of the radio bridge (Condon et al. 1993).
We thus confirm this estimate of the collision time.

\section{Comparison with observations \label{sec:comparison}}

To test the goodness of our ``best-fit'' simulations 19 and 20, we compare the model results to the
following observations of the Taffy galaxy system UGC~12914/15:
\begin{itemize}
\item
stellar distribution,
\item
H{\sc i} and CO gas distributions,
\item
CO velocity field,
\item
H{\sc i} and CO spectra,
\item
polarized radio continuum emission intensity and projected direction of the regular magnetic fields,
\item
radio continuum total power emission.
\end{itemize}
For the comparison we use a unique projection with the primary (UGC~12914) seen at $PA=160^{\circ}$, $i=30^{\circ}$, and an azimuthal viewing angle
(rotation around the angular momentum axis of the primary's disk; see Fig.~9 of Vollmer et al. 2008a) $az=20^{\circ}$.

\subsection{Stellar distribution}

\begin{figure}
  \centering
  \resizebox{7cm}{!}{\includegraphics{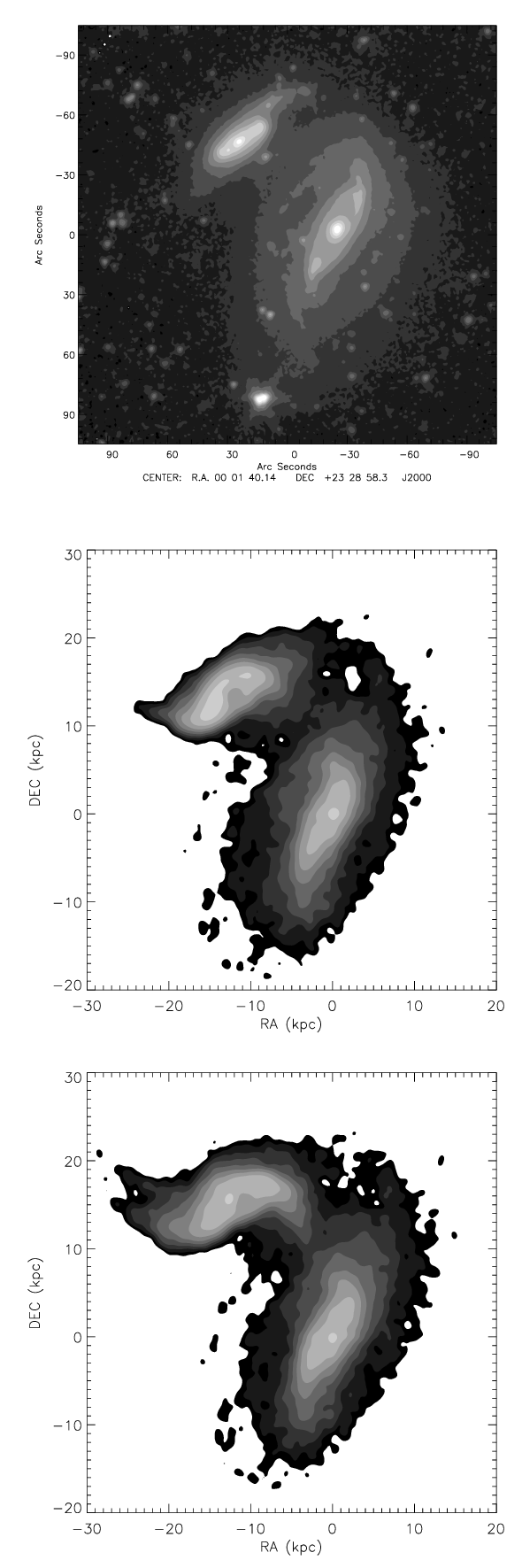}}
  \caption{Stellar surface density distribution. Upper panel: based on Spitzer IRAC $3.6$~$\mu$m observations.
    Middle panel: timestep $t=23$~Myr of simulation 20. Lower panel: timestep $t=23$~Myr of simulation 19. 
    Contour levels are $(1,2,4,8,16,32,64,128,256,512,1024) \times 5$~M$_{\odot}$pc$^{-2}$.
    Projection parameters are $PA=160^{\circ}$, $i=30^{\circ}$, and $az=20^{\circ}$.
  \label{fig:stars}}
\end{figure}

The Spitzer $3.6$~$\mu$m map which traces the stellar mass distribution of the Taffy galaxy system UGC~12914/15 is shown in the
upper panel of Fig.~\ref{fig:stars}.
The smaller galaxy, UGC~12915, has a prominent tidal arm toward the northwest and a faint counterarm to the southeast.
The inner disk of UGC~12914 shows the well-known ring structure at a distance of $\sim 20''=6$~kpc\footnote{We use a distance of $60$~Mpc for
the Taffy galaxy system.} from the center.
Two spiral arms are visible in the north and south of the galactic disk. The stellar surface density of the outer ($6$--$10$~kpc) disk is
higher in the north than in the south. Moreover, UGC~12914 shows a low surface density stellar halo up to a radius
of $\sim 80''=23$~kpc.  
\begin{figure}
  \centering
  \resizebox{7cm}{!}{\includegraphics{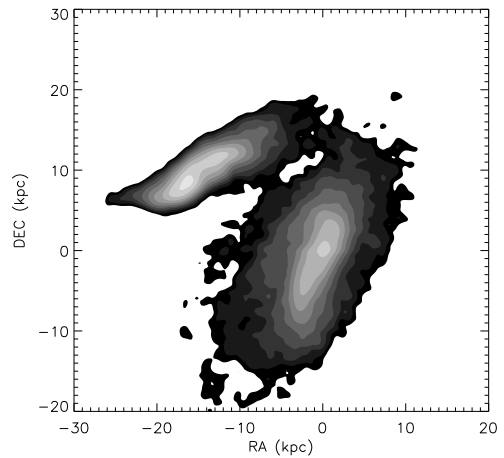}}
  \caption{Stellar surface density distribution of simulation~20 with projection parameters $PA=160^{\circ}$, $i=20^{\circ}$, and $az=0^{\circ}$.
    The stellar morphology of UGC~12915 is better reproduced by the model.
  \label{fig:stars_az0}}
\end{figure}
The model stellar distributions of simulation~19 and 20 are shown in the lower and middle panel of Fig.~\ref{fig:stars}.
There is rough agreement between simulations and observations: all secondary galaxies show a tidal distortion along the major axis. 
However, in both simulations the secondary galaxy is more distorted than observed. Moreover, the curvature of the northern tidal arm 
is stronger than that of UGC~12915 in both simulations. This is caused by the fact that the secondary galaxy is not observed exactly edge-on.
A change of the inclination by $10^{\circ}$ and the azimuthal viewing angle by $20^{\circ}$ of simulation~20 leads to an edge-on projection 
of the secondary galaxy which reproduces the observed stellar distribution of UGC~12915 (Fig.~\ref{fig:stars_az0}). 
We keep the azimuthal viewing angle of
Fig.~\ref{fig:stars}, because the associated projection best reproduces the gas distribution and velocity field.
As already stated in Sect.~\ref{sec:first}, the secondary galaxy of simulation~19 shows more prominent tidal arms than
that of simulation~20 or the observations. Even in the case of an edge-on
projection of the secondary galaxy in simulation~19, the curvature of its northern tidal tail is much stronger than observed.

The resemblance between simulations and observations is better for the primary galaxy, UGC~12914.
Especially the higher stellar surface density in the northern outer stellar disk at ($-6$,$2$)~kpc is reproduced.
This is a direct effect of the passage of UGC~12915 in this region of the disk of UGC~12914.
We conclude that simulation~20 better reproduces the stellar mass distribution of the Taffy galaxy system
UGC~12914/15.

\subsection{Gas distribution}

\begin{figure}
  \centering
  \resizebox{7cm}{!}{\includegraphics{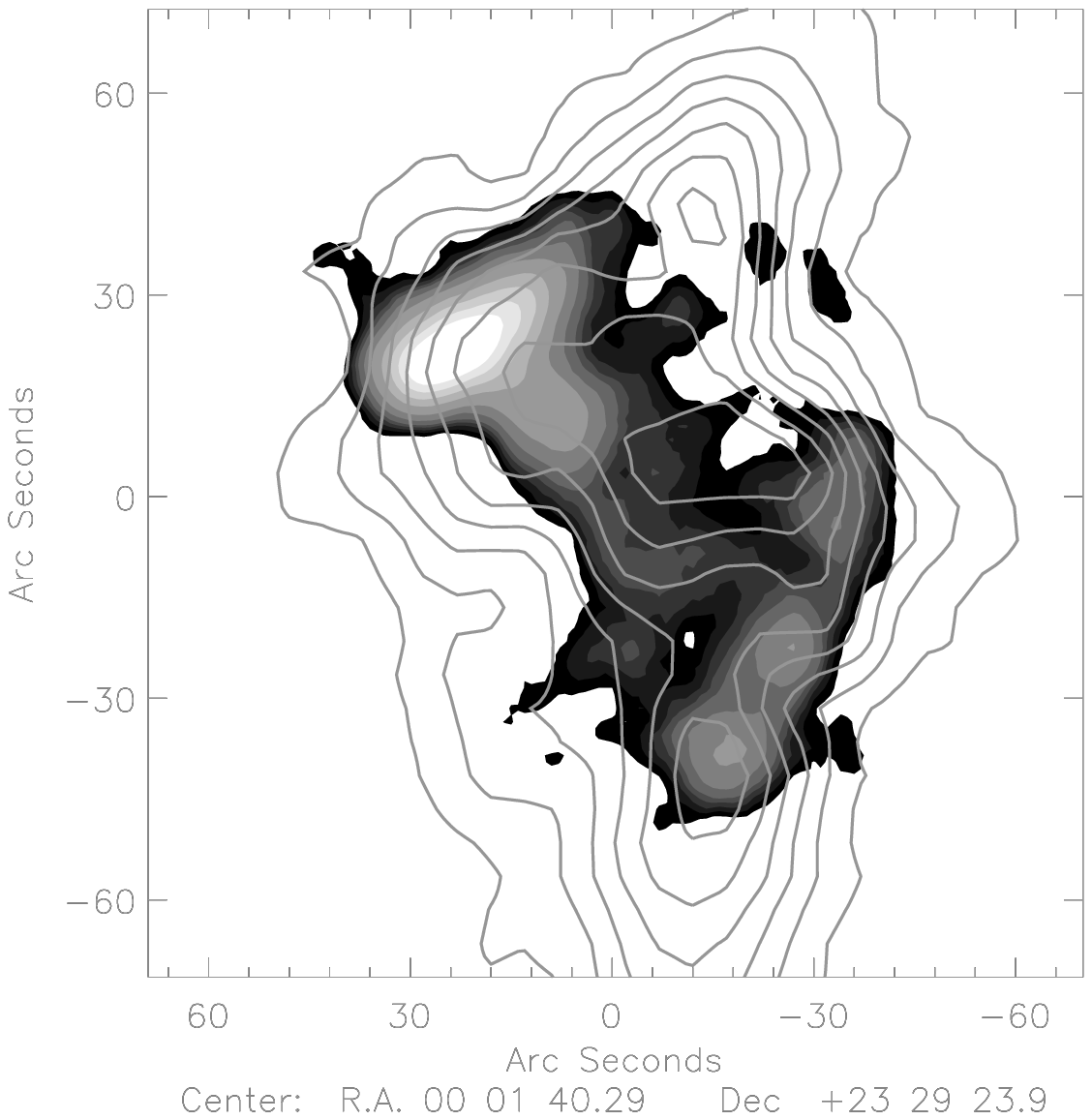}}
  \resizebox{7cm}{!}{\includegraphics{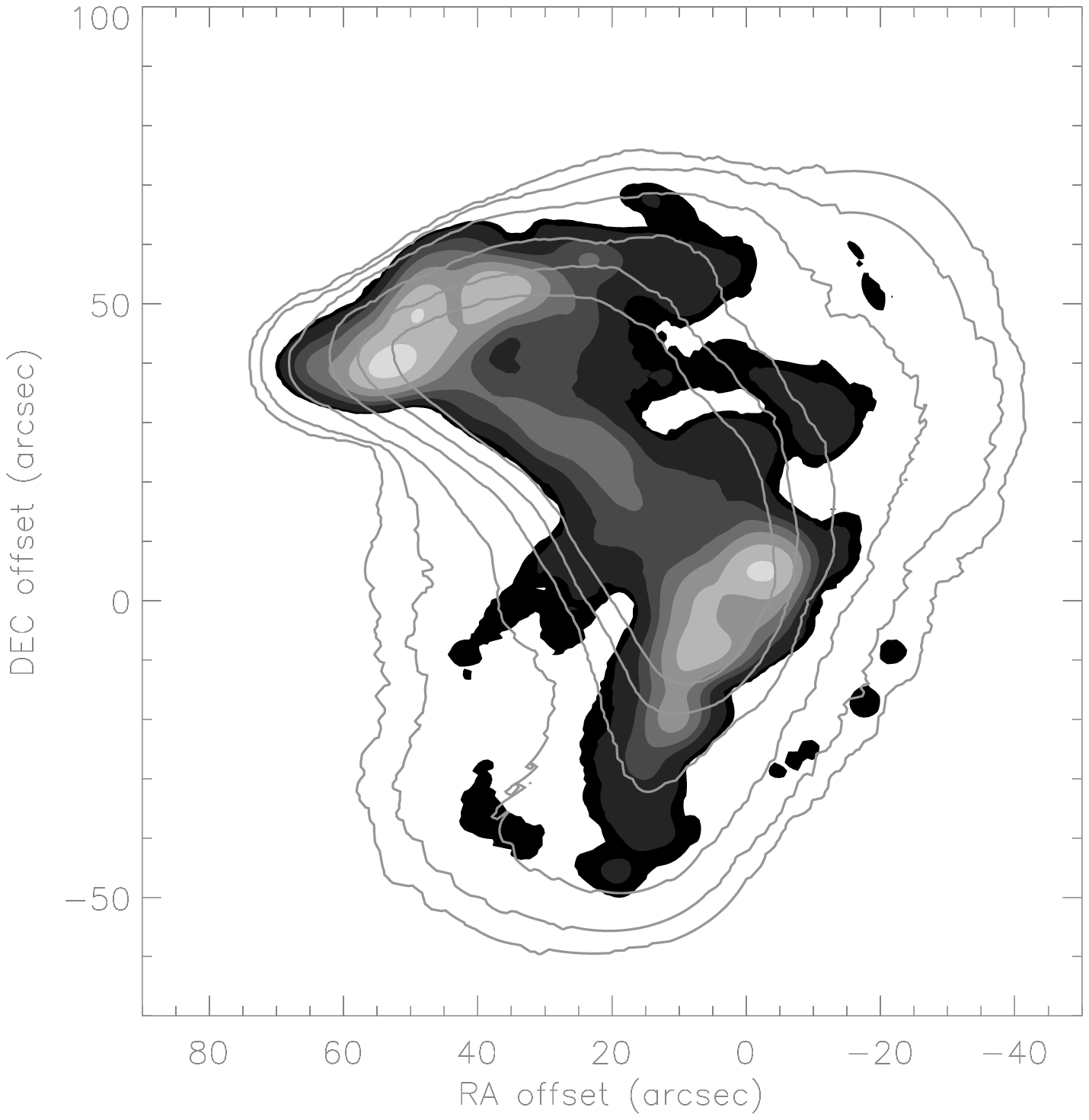}}
  \resizebox{7cm}{!}{\includegraphics{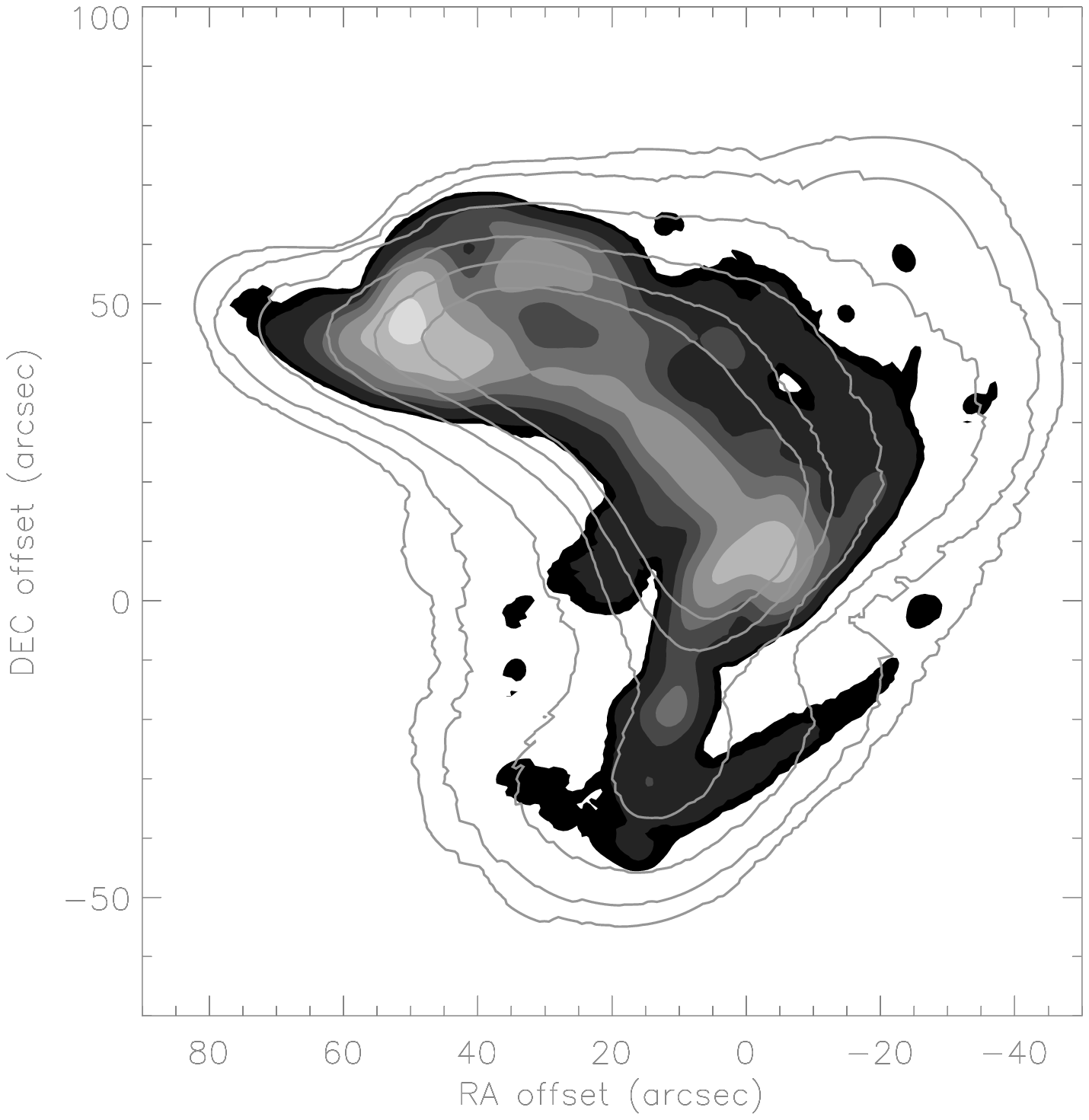}}
  \caption{HI (contour) on CO (greyscale) surface density. Upper panel:
    observations from Condon et al. (1993) and Gao et al. 2003). 
    Contour levels are $(4,8,12,16,20,24,28,32)$~M$_{\odot}$pc$^{-2}$; greyscale levels are
    $(55,60,65,70,75,85,100,125,150,175,200)$~M$_{\odot}$pc$^{-2}$.
    Middle panel: timestep $t=23$~Myr of simulation 20. Contour levels are $(1,2,4,8,12,16,32,64,128)$~M$_{\odot}$pc$^{-2}$;
    greyscale levels are $(1,2,8,16,32,64,128,256)$~M$_{\odot}$pc$^{-2}$.
    Lower panel: timestep $t=23$~Myr of simulation 19. Contour and greyscale levels are the same
    as for the middle panel.
  \label{fig:cohi_model}}
\end{figure}

The model gas clouds represent the ISM as an entity. To distinguish molecular from atomic gas we use a prescription based
on the total gas volume density, which has been successfully applied to the Virgo spiral galaxies NGC~4522 (Vollmer et al. 2008b)
and NGC~4330 (Vollmer et al. 2012a): we assume that the molecular fraction depends linearly on the square root of the
gas density $f_{\rm mol}=M_{\rm mol}/M_{\rm tot}=\sqrt{\rho/(0.03\ {\rm M}_{\odot}{\rm pc}^{-3})}$ and
the molecular fraction cannot exceed unity. A detailed motivation for this prescription is given in Vollmer et al. (2008b).
The atomic and molecular gas surface density distributions are thus given as $\Sigma_{\rm HI}=(1-f_{\rm mol})\,\Sigma_{\rm tot}$
and $\Sigma_{\rm H_{2}}=f_{\rm mol}\,\Sigma_{\rm tot}$. The atomic and molecular gas distributions of the model snapshots of
simulation~19 and 20 are shown together with the CO (Gao et al. 2003) and H{\sc i} (Condon et al. 1993) distributions in Fig.~\ref{fig:cohi_model}.

The outer envelope of the H{\sc i} distribution of model primary galaxies of simulation~19 and 20 are relatively unperturbed.
That of the secondary galaxy follows tidal perturbations visible in the stellar distribution (Fig.~\ref{fig:stars}).
The observed outer H{\sc i} envelope is different, with prominent H{\sc i} maxima to the north and south. In addition, the northern
extension of UGC~12914's gas disk is entirely missing. 
The surface density of the observed H{\sc i} bridge is not constant between UGC~12914 and UGC~12915. 
The maximum of the H{\sc i} bridge is located closer to UGC~12914 and shifted to the northwest with respect
to the line connecting the two galaxy centers. Contrary to observations, the surface density distributions of the model H{\sc i} bridges
are constant along lines connecting the two galaxies. As for the observed H{\sc i} distribution, the regions of
maximum surface density of the model H{\sc i} bridges are shifted to the northwest with respect
to the line connecting the two galaxy centers. This is a direct consequence of the impact which occurs northwest of
the center of the primary galaxy.

The comparison between the model and observed H$_{2}$ distributions is always limited by an uncertain 
conversion factor between CO surface brightness and H$_{2}$ surface density. In the Taffy system, this conversion factor
varies significantly between the disk and bridge regions (Braine et al. 2003, Zhu et al. 2007).
Because of the uncertainty of the local conversion factor, we apply a constant Galactic CO-H$_{2}$ conversion factor
$N({\rm H_{2}})/I_{\rm CO}=10^{20}$~cm$^{-2}$(K\,km\,s$^{-1}$)$^{-1}$
to the whole Taffy system (Zhu et al. 2007). For this comparison one has to bear in mind that the CO-H$_{2}$ conversion factor in the 
bridge might be several times lower than that in the galaxy disks (Braine et al. 2003, Zhu et al. 2007). 

The maximum CO surface brightness and CO luminosity of UGC~12915 are higher than those of UGC~12914. 
The CO emission distributions of UGC~12914 and UGC~12915 are asymmetric along the major axis: they are more extended to the north than to the south.
The maximum CO emission of the bridge is observed close to UGC~12915 and harbors a giant H{\sc ii} region 
(Gao et al. 2003, Braine et al. 2003, Zhu et al. 2007). The projected width of the CO bridge is about half that of the H{\sc i} bridge. 
The CO bridge is offset toward the southeast with respect to the H{\sc i} bridge.

In the model, the H$_{2}$ surface density of the two galactic disks is about the same. 
Whereas the H$_{2}$ surface density of the primary galaxy is more extended to the south in simulation~20, that of the secondary galaxy
is more  extended to the northwest. In simulation~19 the H$_{2}$ distribution of the primary galaxy is more symmetric, and that of the
secondary is strongly distorted by the strong tidal fields. 
The model H$_{2}$ bridges are more continuous than the observed CO bridge. Whereas there is more bridge H$_{2}$ found in the
northeast in simulation~20, the opposite trend is observed in simulation~19. As observed, the widths of the H$_{2}$ bridges 
are about half that of the model H{\sc i} bridges, and they are offset to the southeast with respect to the H{\sc i} bridges.

\subsection{Molecular gas velocity field}

\begin{figure}
  \centering
  \resizebox{\hsize}{!}{\includegraphics{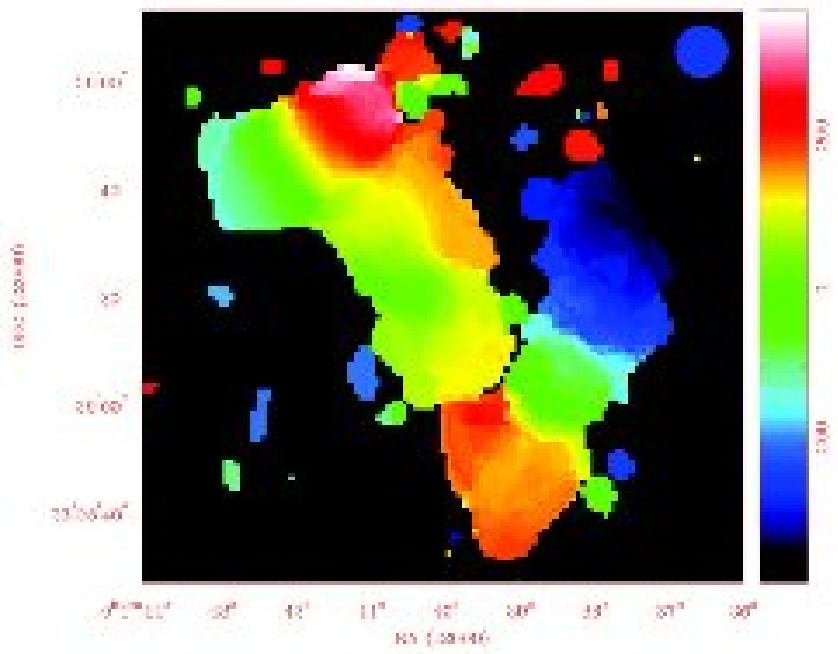}}
  \resizebox{\hsize}{!}{\includegraphics{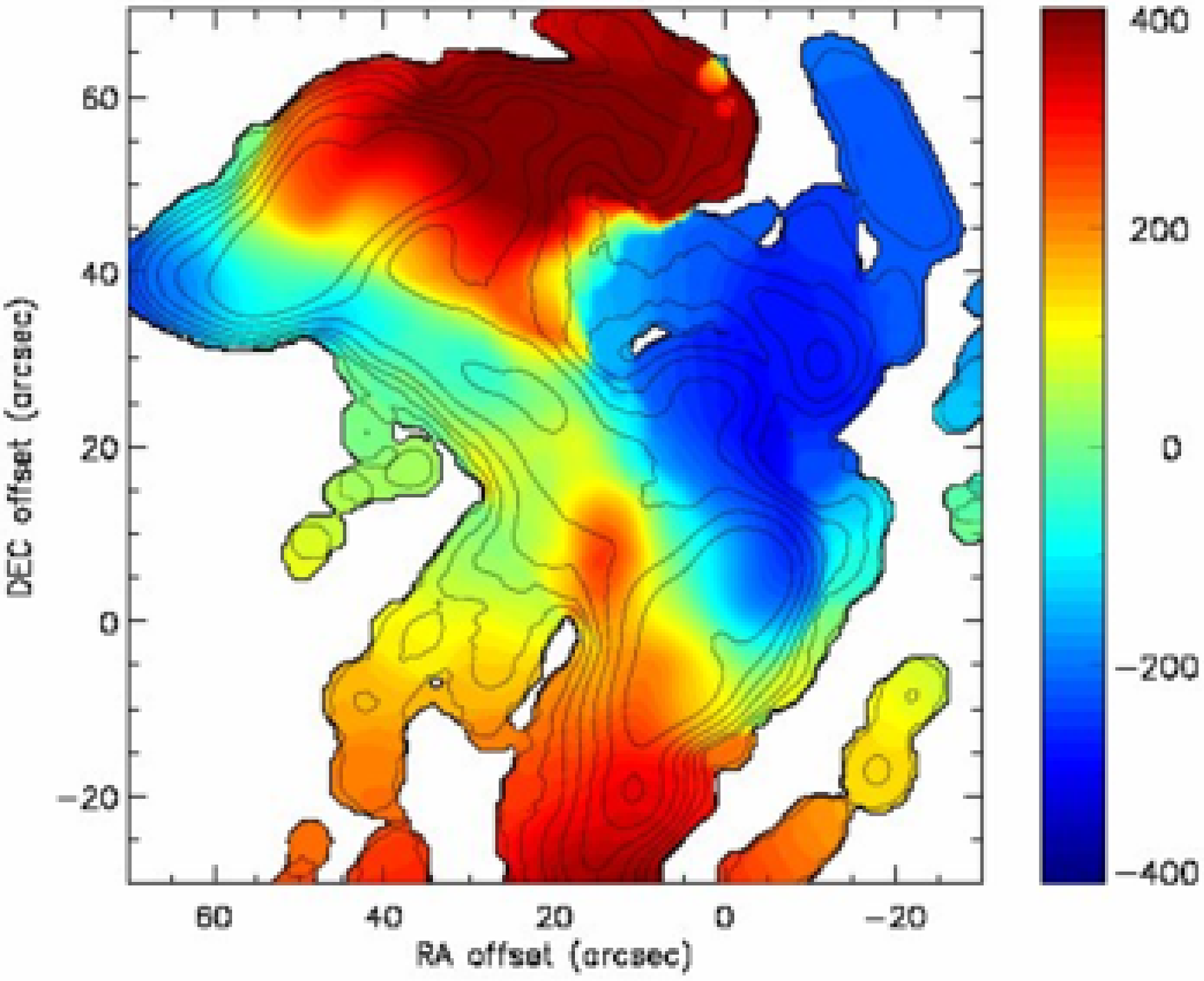}}
  \resizebox{\hsize}{!}{\includegraphics{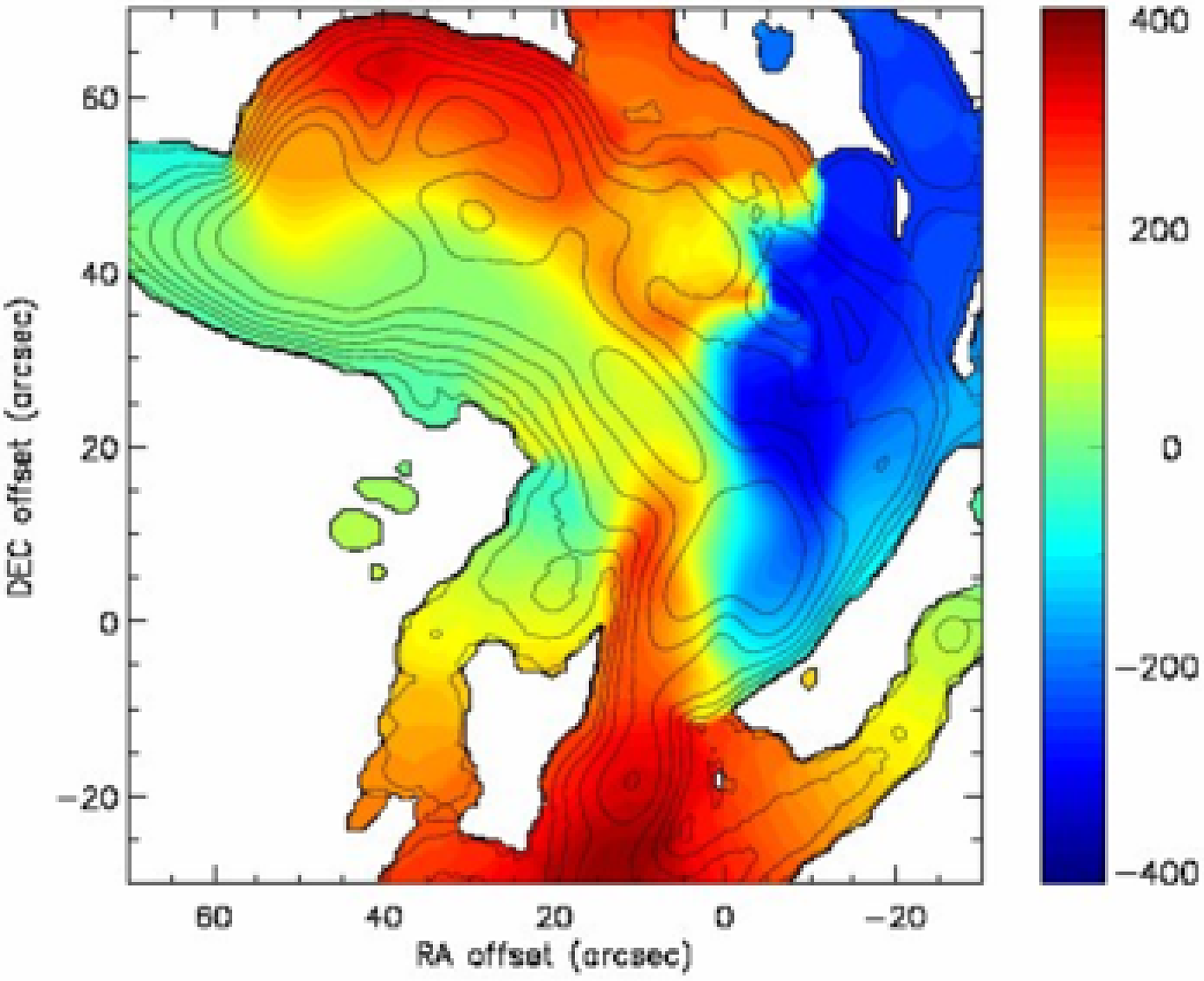}}
  \caption{CO velocity field (colors). Contours: H$_{2}$ surface density from Fig.~\ref{fig:cohi_model}. 
    Upper panel: observations from Gao et al. (2003). Middle panel: timestep $t=23$~Myr of simulation 20. Lower panel:
    timestep $t=23$~Myr of simulation 19.
  \label{fig:cohi_vel_model}}
\end{figure}

The observed CO velocity field of the Taffy system (Gao et al. 2003) is presented in Fig.~\ref{fig:cohi_vel_model} together
with the model molecular gas velocity fields. 
The observed CO velocity field of the galactic disks is quite regular, i.e. the isovelocity contours are approximately
parallel to the disks' minor axes. The molecular gas of most of the bridge region has a zero velocity, i.e. that of UGC~12914.
The velocities become positive ($\sim 150$~km\,s$^{-1}$) in the northwestern bridge region toward UGC~12915 and in the southern bridge region
toward UGC~12914. The velocity gradient of most of the bridge is parallel to that of disk rotation.

Only simulation~20 shows undisturbed disk rotation fields which reproduce the observations. The isovelocity contours
in the disks of simulation~19 are not parallel to the minor axes, but bent, as in barred spiral galaxies.
In both simulations, most of the bridge gas is at zero velocity, in agreement with observations.
The positive velocity of the northern receding side of the secondary galaxy and the southern receding side of the
primary galaxy extend into the bridge region, giving rise to a velocity gradient which is parallel to that of disk rotation, 
again, in agreement with observations. The observed CO velocity structure of the bridge is better
reproduced by simulation~19.

\subsection{H{\sc i} and CO spectra}

The CO observations of Gao et al. (2003) presented in the previous Section were made with the BIMA interferometer.
Due to the lack of very short baselines, the interferometer cannot detect extended structures of molecular gas.
This extended molecular gas component is best detected by a single dish telescope, as the IRAM 30m telescope.
In the presence of complex line shapes, the moment maps contain less information than the individual spectra.
Therefore, we show in Fig.~\ref{fig:img20} 30m CO(1-0) spectra together with VLA H{\sc i} spectra of the Taffy system 
(Fig.~1 of Braine et al. 2003) for the comparison with our simulations.

In the disk regions, where the lines are broad ($\sim 700$~km\,s$^{-1}$), the CO line follows the H{\sc i} line.
In the bridge region the H{\sc i} spectra show a pronounced double-peaked spectrum. Both peaks have comparable widths ($\sim 200$~km\,s$^{-1}$)
and intensities. They are separated by $\sim 100$--$300$~km\,s$^{-1}$. Only the high-velocity part of the H{\sc i} spectra has a CO counterpart. 
On the western side of UGC~12914 a double-line profile is observed in H{\sc i}, with a CO counterpart only
for the low velocity line.
The corresponding model spectra are presented in Fig.~\ref{fig:taffy26new2_az20} for simulation~20 and in 
Fig.~\ref{fig:taffy22new_az20} for simulation~19. The model H{\sc i} and CO lines of the primary galaxy have broad widths 
in both simulations, comparable to the observed line widths. Those of the secondary galaxy are comparable to the observed line widths 
only in simulation~20, because the secondary is seen less edge-on in simulation~19. 
In the bridge region, both simulations display double-line profiles. Both simulations thus reproduce the observed double line profiles.
Moreover, only the low-velocity line has a CO counterpart in the region around $(-7,10)$~kpc for simulation~20 and
around $(-3,10)$~kpc for simulation~19. Although the linewidths of the two components of the observed double-line profiles in the bridge
region are well reproduced by simulation~20, their velocity separation is about twice the observed separation.
On the other hand, simulation~19 reproduces the observed velocity separation between the lines, but the linewidth of the high-velocity
line is about $1.5$ times broader than observed. Both simulations reproduce the observed double-line profile west of the
center of UGC~12914, with a CO line at low velocities and an H{\sc i} line at high velocities.
The causes of the double-line profile will be discussed in Sect.~\ref{sec:collrole}.

\subsection{Radio continuum}

\subsubsection{Observations}

\begin{figure}
  \centering
  \resizebox{\hsize}{!}{\includegraphics{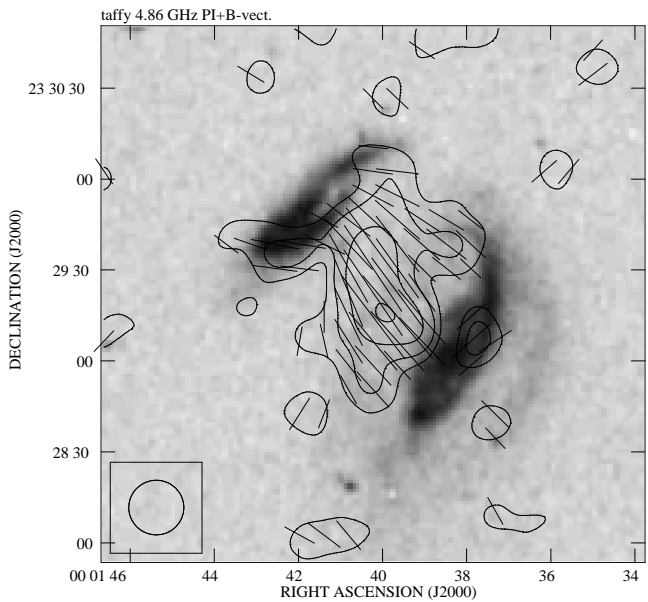}}
  \caption{Observed 6~cm polarized radio continuum contours and projected magnetic field vectors (from Condon et al. 1993)
    on a Digitized Sky survey image of the UGC 12914/5 system. Contour levels are $(3,5,8,12,20,30,50,80,120,200,300,500)
    \times 15~\mu$Jy/beam.
  \label{fig:taffy_pi6c}}
\end{figure}

\begin{figure}
  \centering
  \resizebox{\hsize}{!}{\includegraphics{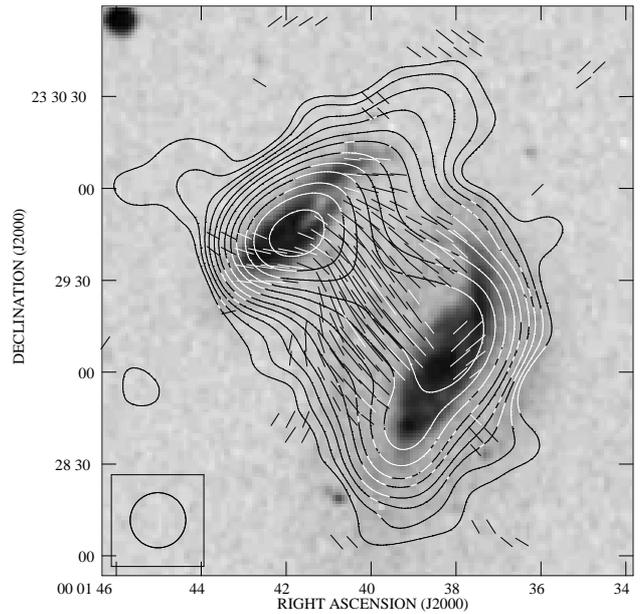}}
  \caption{Observed 6~cm total power radio continuum contours and projected magnetic field vectors (based on
    archival VLA data first published by Condon et al. 1993)
    on a Digitized Sky survey image of the UGC 12914/5 system. Contour levels are $(3,5,8,12,20,30,50,80,120,200,300,500) \times 16~\mu$Jy/beam.
  \label{fig:taffy_pi6tp}}
\end{figure}

We re-reduced archival VLA 6~cm total power and polarized continuum data first published in Condon et al. (1993).
Similar maps of the Taffy system are shown in Drzazga et al. (2011). 
Only the bridge region is detected in polarized emission (Fig.~\ref{fig:taffy_pi6c}). 
The maximum of polarized intensity is located in the southern part of the gas bridge. The projected 
vectors of the large-scale regular magnetic field are parallel to the bridge or perpendicular to the galactic disks.
The total power emission is shown in Fig.~\ref{fig:taffy_pi6tp}. As shown in Condon et al. (1993) the bridge region
shows strong radio continuum emission at 6~cm. Moreover, UGC~12915 has a $\sim 4$ times higher central surface brightness 
than UGC~12914.

\subsubsection{Model}

To calculate the large-scale regular magnetic field for our simulations,
we apply the same procedure as Otmianowska-Mazur \& Vollmer (2003) and Vollmer et al. (2012a).
We solve the induction equation: 
\begin{equation}
{\partial\over\partial t}\vec A=\vec v\times(\nabla\times\vec A)
+\alpha~(\nabla\times\vec A)-\eta~\nabla\times(\nabla\times \vec A)
\label{eq:inductioneq}
\end{equation}
where $\vec{A}=\nabla\times\vec{B}$ is the magnetic potential, $\vec{B}$ the magnetic induction, $\vec{v}$ the large-scale
velocity of the gas, and $\eta$ the coefficient of a turbulent diffusion,
on a 3D grid ($200\times 200 \times 250$). The cell size is $300$~pc.
Time-dependent gas-velocity fields are provided by the 3D dynamical simulations presented in Sect.~\ref{sec:model}.
The induction equation is solved using a second order Godunov scheme with
second order upstream partial derivatives together with a second order
Runge-Kutta scheme for the time evolution.
This results in less numerical diffusion than that of the ZEUS 3D MHD code (Stone \& Norman 1992a,b). 
The 3D velocity field obtained from the N-body code has a discrete distribution. The interpolation to
a regular 3D grid was done with the Kriging method (Soida et al. 2006).
We assume the magnetic field to be partially coupled to the gas via the turbulent diffusion process (Elstner et al. 2000) 
assuming the magnetic diffusion coefficient to be $\eta = 3 \times 10^{25}$~cm$^{2}$s$^{-1}$. 
We do not implement any dynamo process ($\alpha = 0$). 
The initial magnetic field is purely toroidal with an exponential distribution with a scalelength of $1$~kpc in the vertical direction.
Because very rapidly ($\sim 1000$~km\,s$^{-1}$) evolving gas particles lead to numerical artifacts in our code, 
only the primary galaxy, which is kept at rest, is initially magnetized. During the galaxy-galaxy collision the magnetic field 
propagates into the gas bridge and is then amplified by the dynamics of the bridge gas.
Due to the enormous amount of kinetic energy in the bridge gas, the final magnetic field configuration is expected to
be independent of the seed field.

The MHD model does not contain a galactic wind.
The resulting polarized emission is calculated by assuming a density of relativistic electrons that is
proportional to the model gas density $\rho$. This rather crude approximation is motivated by (i) the fact that
in quiescent galaxies, the density of relativistic electrons is approximately proportional to the star formation density
which depends on $\rho^{1{\rm -}1.7}$, and (ii) the transport mechanism of relativistic electrons into the bridge region
during the galaxy collision is unknown.
During the collision, the magnetic field in the bridge region is amplified
by shear and compression motions. The overall morphology of the magnetic field does not
evolve significantly for timesteps around the time of interest ($20$-$35$~Myr).

\begin{figure*}
  \centering
  \resizebox{\hsize}{!}{\includegraphics{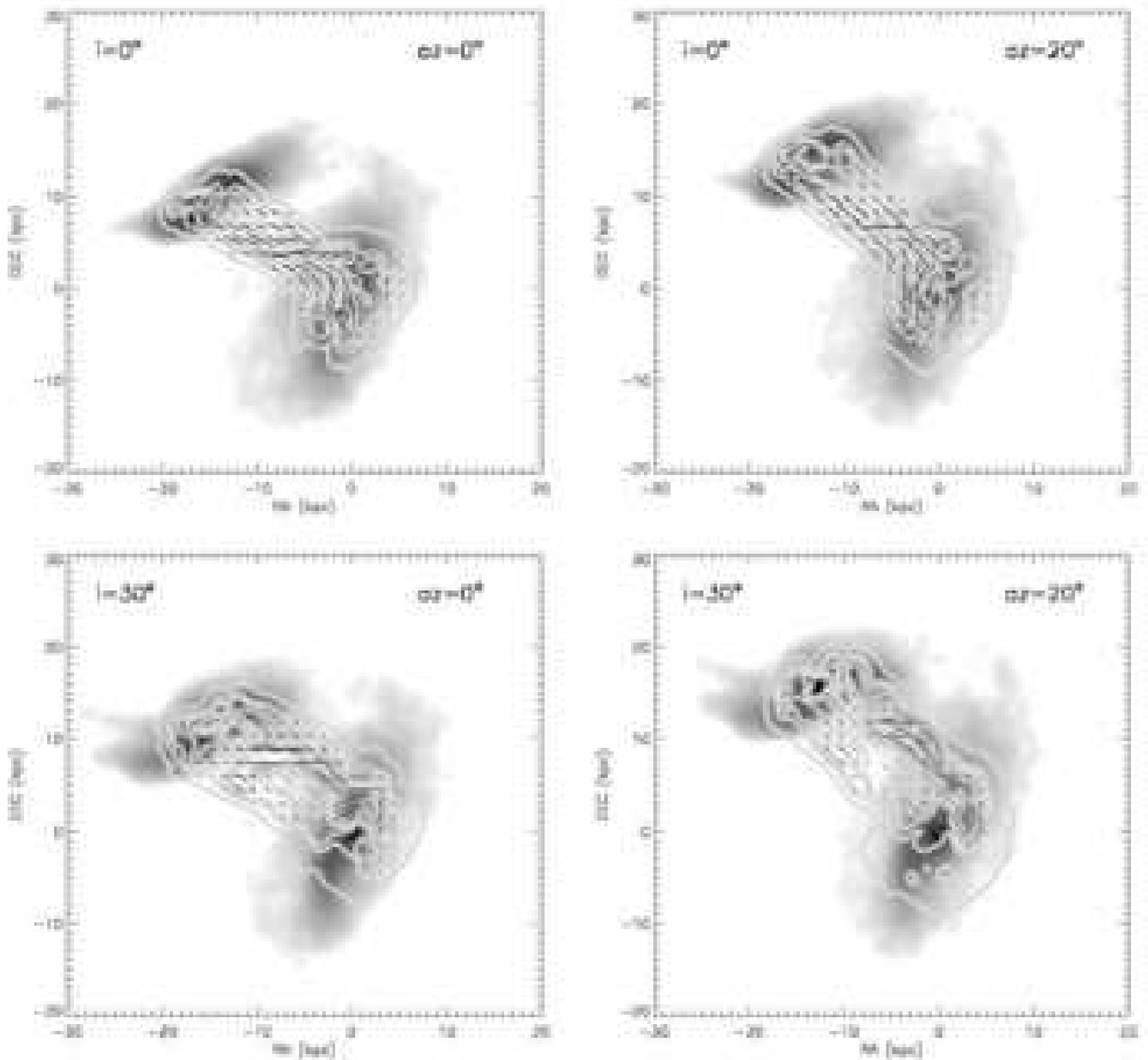}}
  \caption{Model 6~cm polarized radio continuum contours and projected magnetic field vectors on the model 
    stellar surface density distribution for two azimuthal viewing angles.
    Upper panels: simulation 20; lower panels: simulation 19.
  \label{fig:PI+stars_model}}
\end{figure*}

Since the direction of the projected magnetic field vectors strongly depends on projection, we show the
resulting model polarized emission together with the model projected B vectors for simulation~19 and 20
for two azimuthal viewing angles $az=0^{\circ}$ and $20^{\circ}$ in Fig.~\ref{fig:PI+stars_model}. 
In all model polarized emission maps the bridge region is prominent. The model polarized emission of the bridge is as strong as 
that of the two galactic disks. Its maximum is located in the northern/southern part of the bridge in simulation~19/20.
The different azimuthal viewing angles do not lead to significant changes for simulation~20, but do change of the location of the maximum 
polarized emission near the secondary galaxy in simulation~19.
The width of the bridge of polarized emission is $\sim 7$~kpc in simulation~20 and $\sim 12$~kpc in simulation~19.

The model magnetic field vectors are mostly parallel to the bridge
in simulation~20. Only one region close to the primary galaxy displays an offset by $\sim 45^{\circ}$.
In simulation~19 the southwestern part of the bridge has magnetic field vectors parallel to the bridge,
whereas the B vectors of the northeastern part show an offset of $\sim 30^{\circ}$. 
The overall resemblance between the observed and model polarized emission is better for simulation~20 with the
maximum located in the southern part of the bridge and the B vectors parallel to the bridge.

\begin{figure}
  \centering
  \resizebox{9cm}{!}{\includegraphics{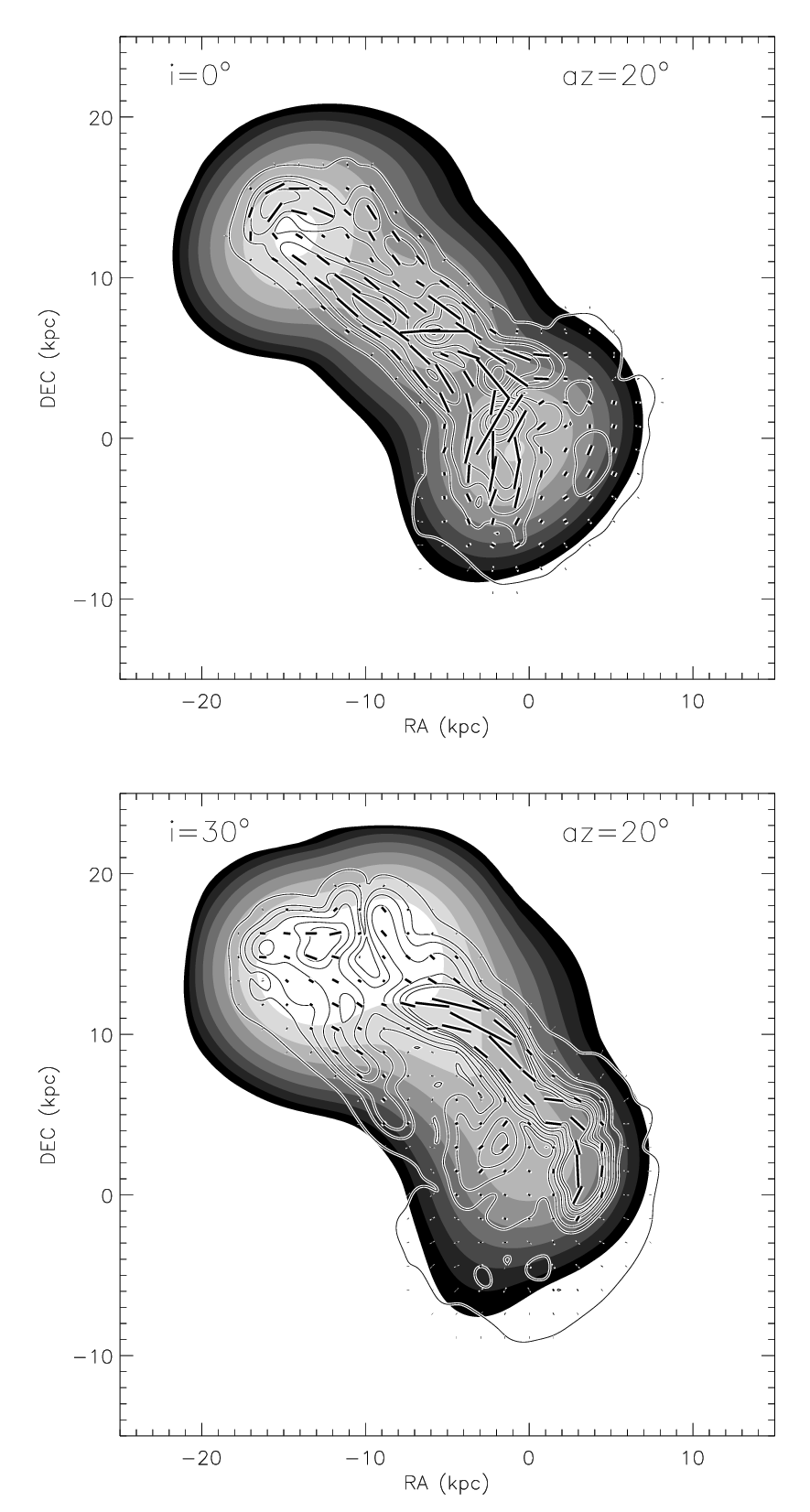}}
  \caption{Model 6~cm total polarized radio continuum emission (contours) and projected magnetic field vectors on the model 
    6~cm total power radio continuum emission distribution (greyscale: $(1,2,4,8,16,32,64,128) \times$ an arbitrary value).
    Upper panel: simulation 20; lower panel: simulation 19.
  \label{fig:PI_TP_model}}
\end{figure}

Whereas the polarized radio continuum emission depends on the large-scale (compared to the beam) regular magnetic field,
the total power emission depends mainly on the small-scale turbulent magnetic field, which dominates the total magnetic field strength.
Since our MHD model only includes the regular magnetic field, we have to derive the turbulent magnetic field in a different way.
For this we assume equipartition between the magnetic field $B$ and the turbulent motions of the gas:
$B^{2}/(8\pi)=1/2 \rho v_{\rm turb}^{2}$, where $\rho$ is the ISM density and $v_{\rm turb}$ its turbulent velocity dispersion.
The total power emission is proportional to the density of relativistic electrons times the square of the turbulent
magnetic field. During the galaxy head-on collision, cosmic ray electrons are dragged into the bridge region.
Since we do not know the mechanisms of the cosmic ray drag, we approximate the total power emission by
the square of the small-scale turbulent magnetic field (Fig.~\ref{fig:PI_TP_model}). In both simulation snapshots the total power
emission of the bridge region is as strong as that of the primary galaxy. This is mainly caused by the high gas velocity
dispersion of the bridge gas (Sect.~\ref{sec:discussion}). The model total power emission of the
secondary galaxy is higher than that of the primary. The model thus qualitatively reproduces the 6~cm total
power emission of the Taffy system (Fig.~\ref{fig:taffy_pi6tp}).

\section{Discussion \label{sec:discussion}}

\subsection{How well does the model reproduce observations?}

To assess the goodness of  the model, we determined the main observational characteristics that a galaxy collision model
should reproduce (Table~\ref{tab:reproduce}). We then checked if these characteristics are reproduced.
In summary, both simulations reproduce the primary's stellar distribution, the prominent H{\sc i} and CO gas bridge,
the offset between the CO and H{\sc i} emission in the bridge, the bridge isovelocity vectors parallel to the bridge,
the H{\sc i} double-line profiles in the bridge region with CO emission only associated to the high-velocity line, 
the large line-widths ($100$--$200$~km\,s$^{-1}$) in the bridge region (simulation 19: $100$--$200$~km\,s$^{-1}$,
simulation 20: $\sim 100$~km\,s$^{-1}$),
the high field strength of the bridge large-scale regular magnetic field, the projected magnetic field vectors parallel to 
the bridge and the strong total power radio continuum emission from the bridge. 
The small separation ($\sim 330$~km\,s$^{-1}$) between the two H{\sc i} lines in the bridge region
is only reproduced by simulation~19 ($\sim 400$~km\,s$^{-1}$), whereas the symmetric velocity field in the galactic 
disks and the southern maximum of polarized intensity in the bridge region are only reproduced by 
simulation~20. The stellar distribution of the secondary galaxy can only be approximately reproduced by simulation~20
with a somewhat different ($20^{\circ}$) projection. The distorted H{\sc i} envelope is reproduced by either simulation.
This is probably due to the collisional description of the model ISM\footnote{However, it cannot be excluded that the initial
gas disks of the colliding galaxies were strongly lopsided.} In the outer parts of the galactic disks, where the 
particle density is low, cloud--cloud collisions become rare. In reality, the H{\sc i} gas of these regions is in form
of filamentary structures with a possibly higher area filling factor than our clouds. The impact of the galaxy collision on this
gas might thus be stronger than predicted by our model. For the gas in the outer galactic disks,
a continuous description of the ISM (SPH or hydro) might thus be preferable.
\begin{table*}
      \caption{Model ability to reproduce observational characteristics .}
         \label{tab:reproduce}
      \[
         \begin{tabular}{lll}
           \hline
           \noalign{\smallskip}
             & simulation~19 & simulation~20 \\ 
	   \hline
	   \hline
	   \noalign{\smallskip}
	   primary's stellar distribution & yes & yes \\
	   \noalign{\smallskip}
	   \hline
	   \noalign{\smallskip}
	   secondary's stellar distribution & too distorted & OK with different projection ($20^{\circ}$) \\
	   \noalign{\smallskip}
	   \hline
	   \noalign{\smallskip}
	   distorted H{\sc i} envelop & no & no \\
	   \noalign{\smallskip}
	   \hline
	   \noalign{\smallskip}
	   prominent H{\sc i} and CO gas bridge & yes & yes \\
	   \noalign{\smallskip}
	   \hline
	   \noalign{\smallskip}
	   CO bridge emission offset w.r.t. H{\sc i} to the south & yes & yes \\
	   \noalign{\smallskip}
	   \hline
	   \noalign{\smallskip}
	   symmetric gas velocity fields in the disks & no & yes \\
	   \noalign{\smallskip}
	   \hline
	   \noalign{\smallskip}
	   isovelocity contours parallel to the bridge & yes & yes \\
	   \noalign{\smallskip}
	   \hline
	   \noalign{\smallskip}
	   H{\sc i} double-line profile in the bridge region & yes & yes \\
	   \noalign{\smallskip}
	   \hline
	   \noalign{\smallskip}
	   CO emission only associated with the high-velocity line & yes & yes \\
	   \noalign{\smallskip}
	   \hline
	   \noalign{\smallskip}
	   large gas linewidths in the bridge region & & \\
	    ($100$--$200$~km\,s$^{-1}$) & yes & yes \\
	   \noalign{\smallskip}
	   \hline
	   \noalign{\smallskip}
	   small separation between the double lines  & & \\
	   ($\sim 330$~km\,s$^{-1}$) & yes & no \\
	   \noalign{\smallskip}
	   \hline
	   \noalign{\smallskip}
	   high field strength of the bridge regular & & \\
	   magnetic field & yes & yes \\
	   \noalign{\smallskip}
	   \hline
	   \noalign{\smallskip}
	   projected B field vectors parallel to the bridge & with offset of $\sim 30^{\circ}$ & yes \\
	   \noalign{\smallskip}
	   \hline
	   \noalign{\smallskip}
	   polarized intensity maximum offset to the south & & \\
	   of the bridge & no & yes \\
	   \noalign{\smallskip}
	   \hline
	   \noalign{\smallskip}
	   gas bridge strong in total power radio & & \\
	   continuum emission & due to high turbulent velocity & due to high turbulent velocity \\
	   \noalign{\smallskip}
	   \hline
        \end{tabular}
      \]
\end{table*}
The present simulations thus reproduce qualitatively most of the observed multiwavelength characteristics.
Simulation~20 better reproduces observations than simulation~19.
Despite the calculation of $45$ models, we were not able to find one single initial condition and projection that reproduces all 
observed characteristics.
We thus decided to stop our parameter study. We are confident that the actual interaction parameters of the Taffy system
UGC~12914/15 are not very far from those used for simulation~19 and 20 (Table~\ref{tab:simulations1}).

\subsection{The role of collisions \label{sec:collrole}}

For simulation~19 where the secondary's disk is initially inclined by $30^{\circ}$ with respect to the primary's disk,
the disk of the secondary galaxy is strongly distorted by the tidal fields (Fig.~\ref{fig:stars}).
The secondary's tidal distortion is much less strong in simulation~20, where the two disks are initially parallel.

What is the role of tidal distortions and distortions caused by the collisional nature of the modelled ISM?
To investigate the role of the cloud--cloud collisions, we re-simulated simulation~19 suppressing the cloud collisions.
It is surprising that there are regions in the bridge that show a double-line profile reminiscent of the
observed double-line profile (Fig.~\ref{fig:img20}) with comparable linewidths and line separations.
Contrary to observations, the low-velocity line of the model is most prominent in CO. Moreover, the model lines in the bridge
region are much weaker than the observed ones. The double-lines in this simulation are produced by the projection
of two spatially different gas flows. Tidal effects thus can create double-line profiles, but they are
not the main mechanism for the creation of the observed double-line profiles in the bridge of the Taffy system UGC~12914/15.

\subsection{Isolating the bridge region}

There is a debate in the literature (Gao et al. 2003, Braine et al. 2003, Zhu et al. 2007) about how much gas is located
in the bridge between UGC~12914 and UGC~12915. This debate has three aspects: (i) the delimitation of the bridge region,
(ii) the CO--H$_{2}$ conversion factor of the bridge gas, and (iii) the H{\sc i} mass belonging to the bridge.
Our model can help to clarify points (i) and (iii). For this purpose we have separated the bridge region in 3D from the disk regions
as described in Sec.~\ref{sec:simulations}: we define the bridge region as
a vertical cylinder with an infinite radius extending from $3$~kpc to $12$~kpc with respect to the disk plane of the 
primary galaxy. The model spectra of simulation~19 showing only the bridge region is presented in Fig.~\ref{fig:taffy22new_az20_bridge}.
In the region of the secondary's stellar tidal tail a double-line profile is found with a dominant high-velocity component.
The linewidth of this component is large ($\sim 200$~km\,s$^{-1}$). The 3D bridge region extends from the eastern border of
the primary's to the western border of secondary's high surface brightness disk. Almost all lines from the bridge
region are strong in CO emission. Since our model H{\sc i}/CO separation is based on gas density, this means that
the bridge gas has a relatively high gas density.

We thus conclude that the bridge region delimited by Braine et al. (2003, Fig.~\ref{fig:img20}) is somewhat too small.
It should extend until the border of the high surface brightness disk of UGC~12914. Within the disk, only the
high-velocity component of the double-lines belongs to the bridge. 
Based on these conclusions we revise the bridge H{\sc i} mass to be $M_{\rm HI bridge} \sim 3 \times 10^{9}$~M$_{\odot}$.
Applying a CO--H$_{2}$ conversion factor of $N({\rm H_{2}})/I_{\rm CO}=4 \times 10^{19}$~cm$^{-2}$(K\,km\,s$^{-1}$)$^{-1}$
(Zhu et al. 2007), the mass of the molecular gas located in the bridge is also $M_{\rm H_{2} bridge} \sim 2 \times 10^{9}$~M$_{\odot}$.
The recently discovered warm H$_{2}$ (Peterson et al. 2012) might add up to $10^{9}$~M$_{\odot}$ to the molecular gas content if
the H$_{2}$-emitting gas is different from the CO-emitting gas
\footnote{It is very likely that at least some of the gas detected in warm H$_2$ emission by Peterson et al. (2012)
is the same as what we detect in CO. While the warm H$_2$ they detect is very different from what is seen in galactic giant
molecular clouds, ($i$) the $^{13}$CO emission is very weak in the bridge (Braine et al. 2003), such that even the $^{12}$CO emission may 
be optically thin and ($ii$) the CO(2--1)/CO(1--0) ratio is compatible with warm optically thin emission from low-density molecular gas.  
It would be very useful to have observations of higher$-J$ transitions in order to break this degeneracy.}.
The bridge gas thus has a molecular fraction of $M_{\rm H_{2}}/M_{\rm HI} \sim 1$.

\subsection{Cloud collisions in the gas bridge}

\begin{figure}
  \centering
  \resizebox{\hsize}{!}{\includegraphics{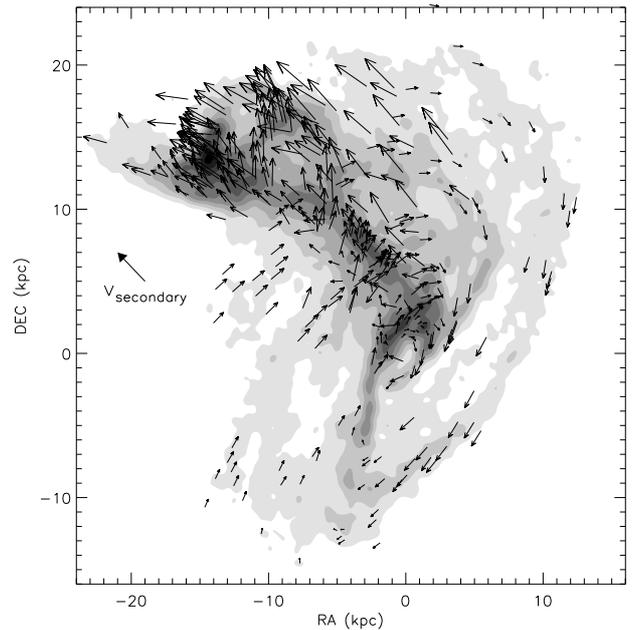}}
  \caption{Transverse gas velocities of simulation 19 on the total gas surface density distribution. The primary galaxy is at rest.
    The thick arrow indicates the transverse velocity of the secondary galaxy.
  \label{fig:velfield}}
\end{figure}

The bridge gas has a high surface density ($\Sigma_{\rm HI} \sim 30$~M$_{\odot}$pc$^{-2}$; $\Sigma_{\rm CO} \sim 50$~M$_{\odot}$pc$^{-2}$;
see Fig.~\ref{fig:cohi_model}). The bright CO emission of the disk gas implies high gas densities ($\sim 1000$~cm$^{-3}$) of a significant 
portion of the bridge gas. Our model reproduces these findings. However, there is no star formation associated with most of the bridge gas
(Gao et al. 2003, Braine et al. 2003, Zhu et al. 2007). Only the bridge gas of very high surface density close to UGC~12915
is vigorously forming stars.
What makes the bridge gas not form stars despite its high density? To investigate this question, we need to have a closer look
at the physical properties of the bridge gas. Fig.~\ref{fig:velfield} shows the gas velocities in the plane of the sky.
We identify large-scale colliding flows from the south of the bridge into the bridge region of highest surface density around the position
$(-3,7)$~kpc. These flows lead to a compression of the gas and enhances the cloud--cloud collision rate.

\begin{figure*}
  \centering
  \resizebox{\hsize}{!}{\includegraphics{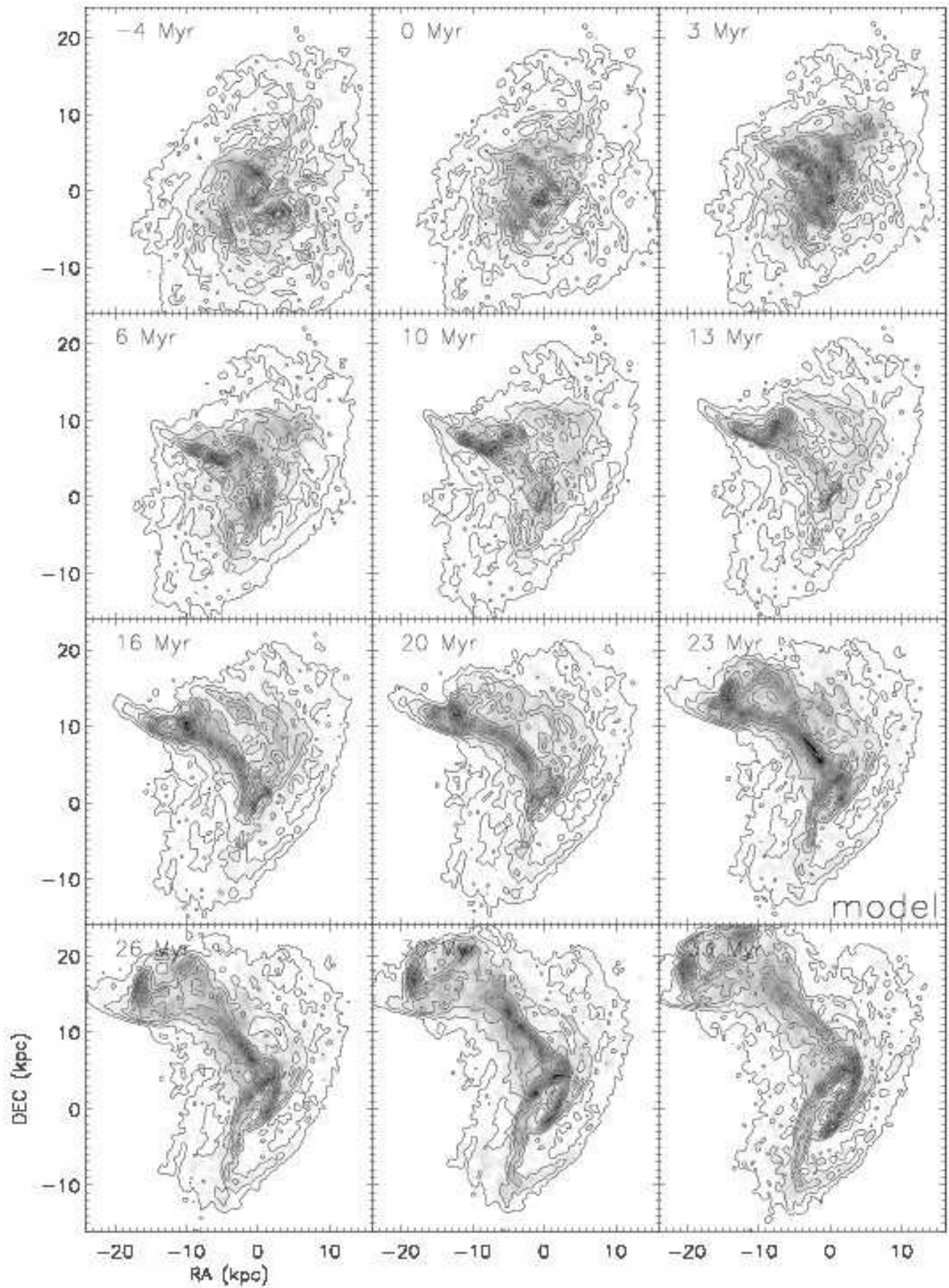}}
  \caption{Simulation 19. Contours of the total gas surface density on the distribution of cloud-cloud collisions (greyscale).
    The timestep $t=0$~Myr corresponds to the impact. 
  \label{fig:collisions+gas-evol}}
\end{figure*}

The time evolution of the local cloud--cloud collision rate of simulation~19 is presented in Fig.~\ref{fig:collisions+gas-evol}.
As expected, the local cloud collision rate generally follows the gas surface density. Before the galaxy collision most cloud--cloud 
collisions occur in the inner galactic disks and in spiral arms where the gas density is high. This situation does not change 
significantly until $\sim 10$~Myr after the impact. The local cloud collision rate in the bridge region is as high as in
the spiral arms of the unperturbed disks. After $\sim 16$~Myr the local collision rate in the portion of the bridge 
near the primary galaxy begins to increase. It reaches its maximum $\sim 23$~Myr after the impact.
This corresponds exactly to the timestep of interest. For $t > 23$~Myr the local cloud collision rate in the bridge decreases again. 
The bridge region at the time of interest is thus in a phase of active compression with a high input of mechanical energy.
For $t > 23$~Myr ring structures form in both galactic disk where the local cloud collision rate is enhanced.

\subsection{Gas densities and velocity dispersions in the bridge}

\begin{figure*}
  \centering
  \resizebox{\hsize}{!}{\includegraphics{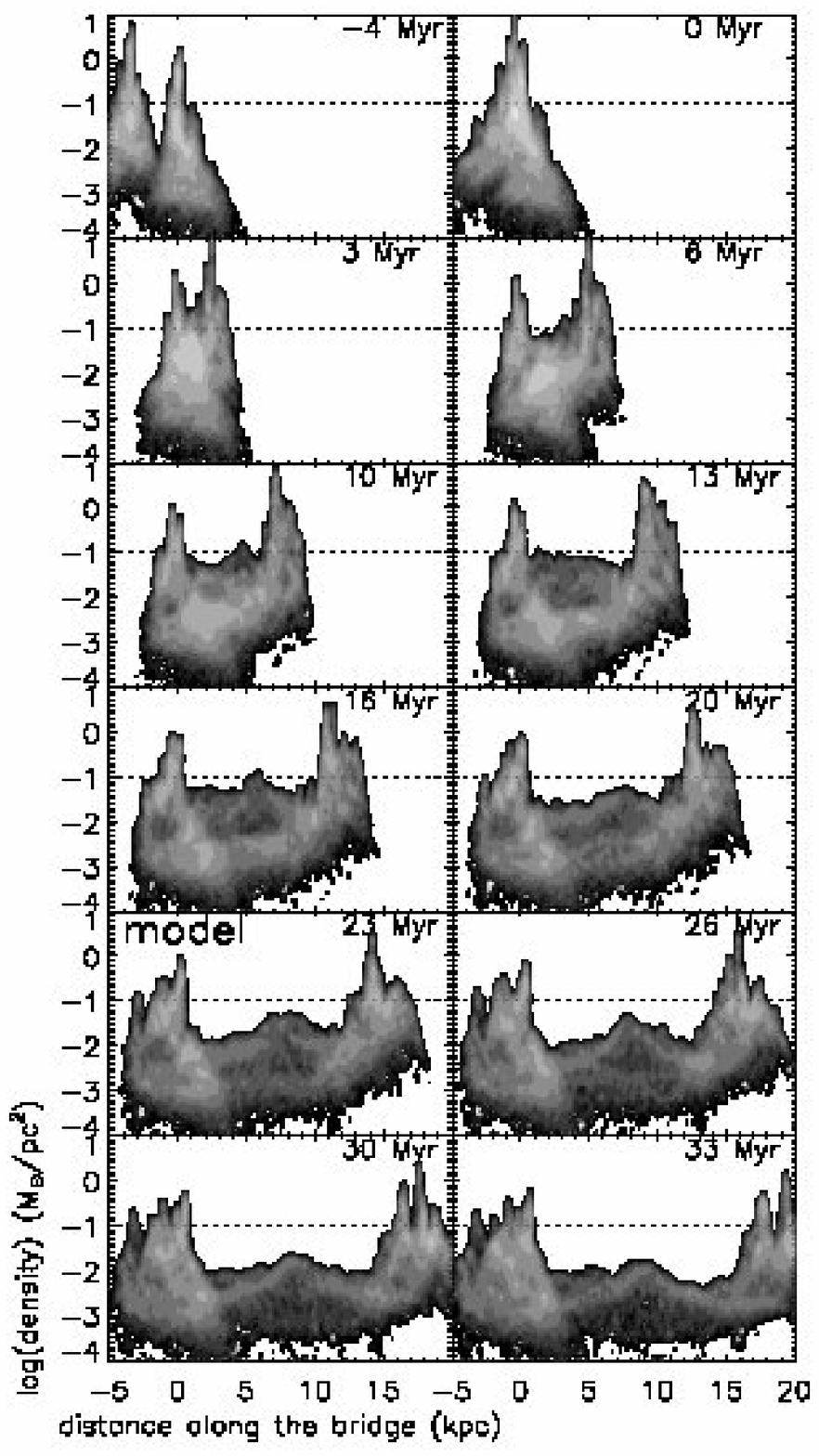}\includegraphics{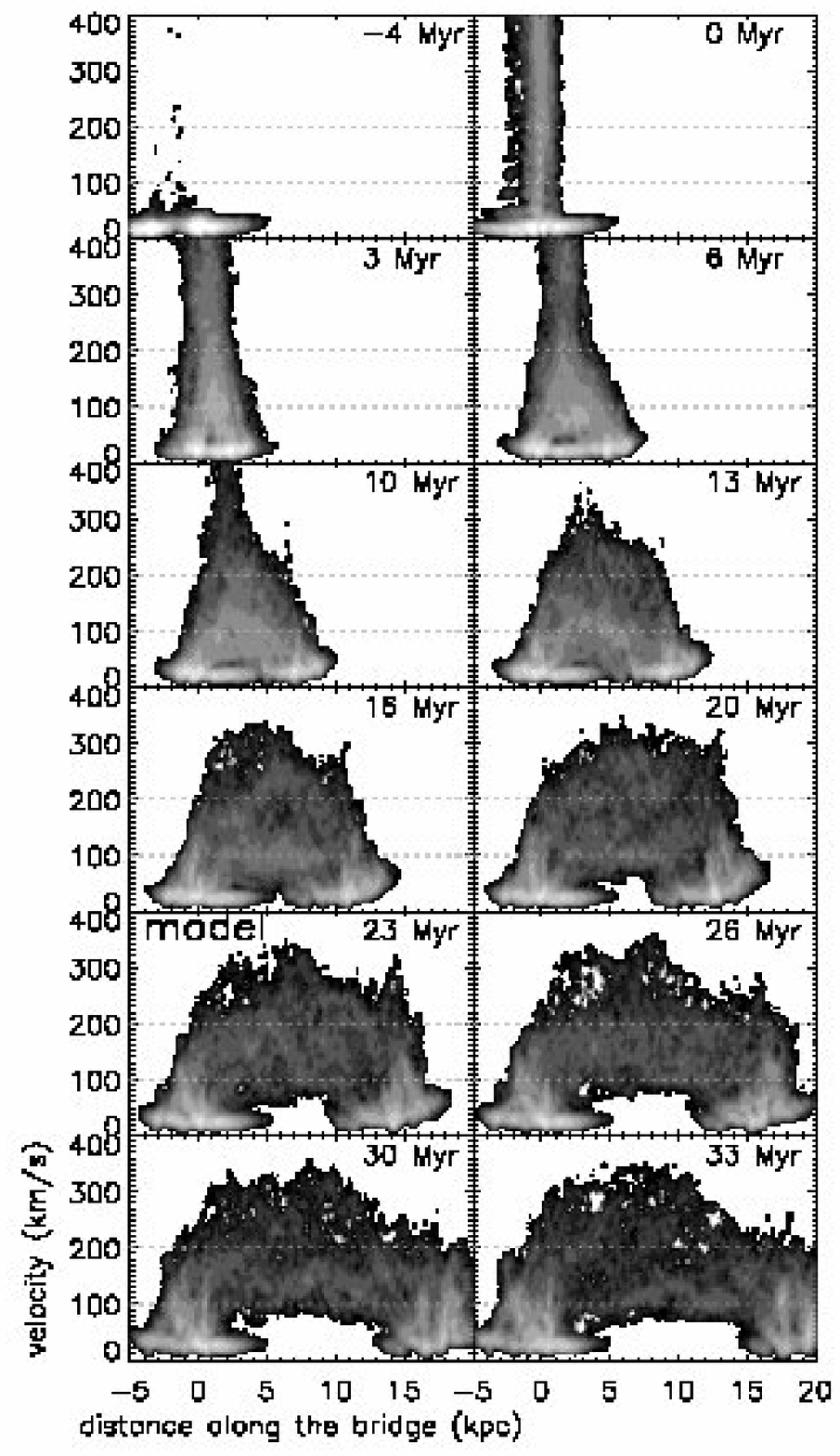}}
  \caption{Evolution of the bridge gas properties of simulation 20. Left panels: total gas volume density as a function of distance
    along the bridge. Right panels: 3D velocity dispersion as a function of distance along the bridge.
  \label{fig:dist_TAFFY26new2}}
\end{figure*}

\begin{figure*}
  \centering
  \resizebox{\hsize}{!}{\includegraphics{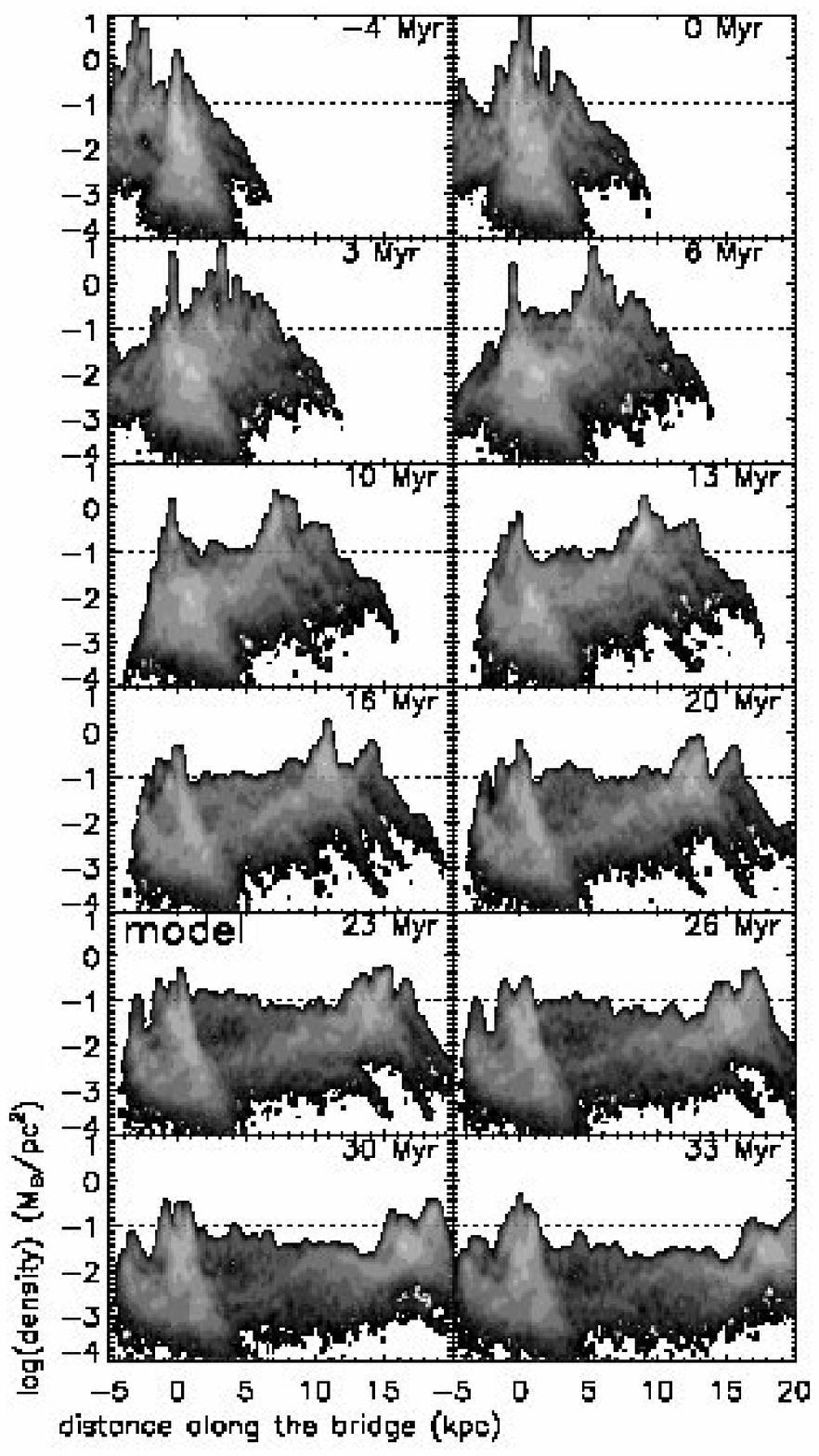}\includegraphics{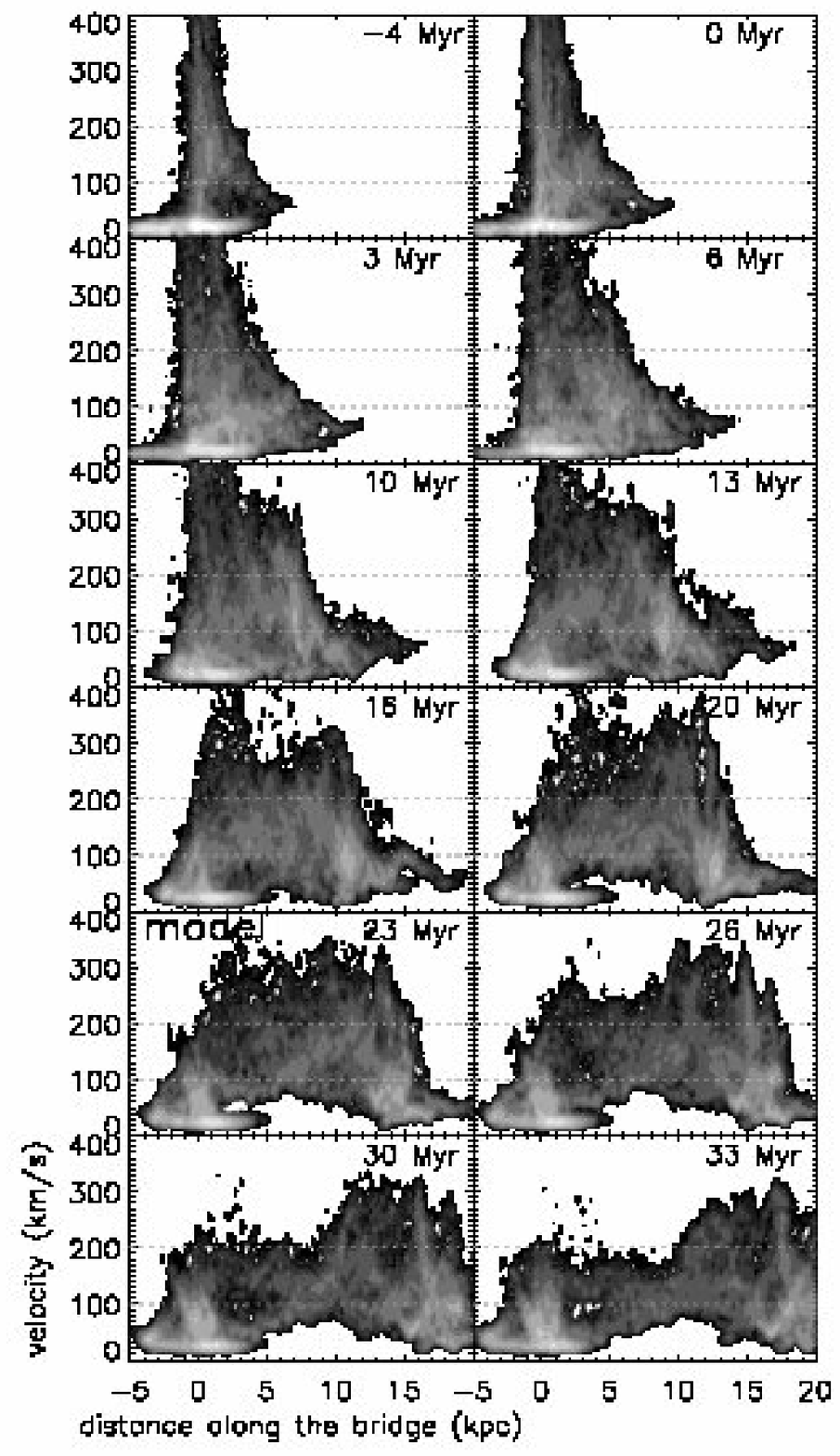}}
  \caption{Evolution of the bridge gas properties of simulation 19. Left panels: total gas volume density as a function of distance
    along the bridge. Right panels: 3D velocity dispersion as a function of distance along the bridge.
  \label{fig:dist_TAFFY22new}}
\end{figure*}

To further investigate the physical properties of the disk region, we present the time evolution of the gas density and
3D velocity dispersion of simulation~20/19 in Figs.~\ref{fig:dist_TAFFY26new2}/\ref{fig:dist_TAFFY22new}.
With our kpc resolution, the densities of the unperturbed disk range between $10^{-3}$ and $1$~M$_{\odot}$pc$^{-3}$, the velocity dispersions
between ($10$ and $30$~km\,s$^{-1}$). Due to the galaxy collision, the velocity dispersion increases up to
$400$~km\,s$^{-1}$. The velocity dispersion of most of the gas located in the bridge is between
$100$ and $200$~km\,s$^{-1}$. In addition, there are regions in the bridge where the velocity dispersion reaches 
$200$ to $300$~km\,s$^{-1}$. In simulation~19 these regions are mostly close to the secondary galaxy for $t > 23$~Myr. 
In both simulations, the mean velocity dispersion of the bridge gas decreases for $t > 23$~Myr.
In simulation~20, the velocity dispersion is about constant along the bridge. On the other hand, in simulation~19
the bridge velocity dispersion increases slowly from the primary to the secondary galaxy along the bridge.
At $t=30$~Myr an X-structure forms. 

In both simulations the bridge contains a high-density filament embedded into low-density regions. The density distribution is
not continuous between these extremes. The density of the high-density filament at the time of interest is of the order of 
$10^{-2}$~M$_{\odot}$pc$^{-3}$ in both simulations. Due to the expansion of the bridge region with time, the overall gas density
in the bridge decreases for $t > 20$~Myr in simulation~20 and for  $t > 26$~Myr in simulation~19.
Whereas the high-density gas has a velocity dispersion higher than $100$~km\,s$^{-1}$, the low-density gas has a velocity dispersion
lower than $100$~km\,s$^{-1}$ in both simulations.
We identify the high-density ($10^{-2}$~M$_{\odot}$pc$^{-3}$), high-velocity-dispersion ($> 100$~km\,s$^{-1}$) bridge regions
with those regions emitting in the CO line. The observed large linewidth ($\sim 200$~km\,s$^{-1}$) of the CO lines in the
bridge are thus consistent with a high intrinsic velocity dispersion of the bridge gas in our model.

Assuming equipartition between the kinetic and the magnetic energy density $\rho v_{\rm disp}^2 = B^2/(8\pi)$,
a velocity dispersion of $v_{\rm disp}=150$~km\,s$^{-1}$, and a total gas density of $\rho=0.01$~M$_{\odot}$pc$^{-3}$, we estimate the total 
magnetic field strength in the high-density bridge region to be $B \sim 65$~$\mu$G.
The total magnetic field strength in the low-density bridge region ($\rho=10^{-3}$~M$_{\odot}$pc$^{-3}$, $v_{\rm disp}=50$~km\,s$^{-1}$)
is  $B \sim 7$~$\mu$G. The latter value is close to the magnetic field strength derived from the $1.4$~GHz specific intensity
assuming equipartition between the energy of the relativistic electrons and the magnetic field (Condon et al. 1993).
The lifetime of synchrotron-emitting electrons is $t \sim 1\,(B/1\mu{\rm G})^{-1.5}(\nu/1{\rm GHz})^{-0.5}$~Gyr, where $\nu$ is the frequency of the
observation. This lifetime is $\sim 1$~Myr for the high-density regions and $\sim 25$~Myr for the low-density regions at $\nu=4.8$~GHz.
In the absence of newly injected relativistic electrons produced in supernova explosions, the radio continuum emission of the
high-density regions drops under the detection limit after a few Myr. That of the low-density regions persists for several $10$~Myr.

In the bridge region, star formation only occurs in the giant H{\sc ii} region close to UGC~12915. Most of the bridge region is devoid
of star formation and thus devoid of the injection of relativistic electrons. 
In this latter region, the high surface brightness $20$~cm emission follows the H{\sc i} and not the CO distribution
(Fig.~8 of Gao et al. 2003), i.e. it follows the low-density atomic gas.
Based on our model, we thus conclude that the observed synchrotron radio continuum emission of the bridge region outside the
giant H{\sc ii} region near UGC~12915 mainly stems from the low-density, high-velocity-dispersion bridge gas which is observed in H{\sc i}.

The high mechanical energy input into the bridge region is consistent with the recent detection of strong, resolved emission from warm H$_{2}$ in
the bridge of the Taffy galaxy system UGC~12914/15 (Peterson et al. 2012). The latter authors state that ``since the cooling time of
warm H$_{2}$ is short ($\sim 5000$~yr), shocks must be permeating the molecular gas bridge region in order to continue heating the H$_{2}$''.
Based on our model, we identify the energy source to be the mechanical energy from compressive colliding gas flows in the bridge region.
Peterson et al. (2012) estimate the mean surface brightness of warm H$_{2}$ emission to be 
$\Delta E/(\Delta A \Delta t) \sim 1.7 \times 10^{-7}$~W\,m$^{-2}$. The mechanical energy input is 
$\Delta E/(\Delta A \Delta t) \sim \rho v_{\rm disp}^{3}$. For the high-density bridge gas 
($\rho=0.01$~M$_{\odot}$pc$^{-3}$, $v_{\rm disp}=150$~km\,s$^{-1}$) we obtain 
$\Delta E/(\Delta A \Delta t) \sim 2.5 \times 10^{-6}$~W\,m$^{-2}$, for the low-density gas 
($\rho=10^{-3}$~M$_{\odot}$pc$^{-3}$, $v_{\rm disp}=50$~km\,s$^{-1}$) $\Delta E/(\Delta A \Delta t) \sim 10^{-8}$~W\,m$^{-2}$.
We thus conclude that only the high-density gas undergoes a high enough mechanical energy input to produce the observed
emission of warm H$_{2}$. Our model suggests that we observe this galaxy head-on collision near the time of maximum CO and H$_{2}$ emission.
The physical properties of the bridge gas are similar to those of the dense gas in the large-scale shock of Stephan's Quintet.
There, the molecular gas carries a large fraction of the gas kinetic energy involved in the collision which has not been thermalized
yet (Guillard et al. 2012).

\subsection{The low star formation efficiency in the bridge}

The local star formation rate was calculated from the FUV luminosities corrected by the $24$~$\mu$m IR emission following Leroy et al. (2008).
This method takes into account the UV photons from young massive stars which escape the galaxy and those which
are absorbed by dust and re-radiated in the far infrared:
\begin{equation}
\dot{\Sigma}_{*} = 8.1 \times 10^{-2}\ I({\rm FUV}) + 3.2 \times 10^{-3}\ I({\rm 24\mu m})\ ,
\end{equation}
where $I({\rm FUV})$ is the GALEX far ultraviolet and $I({\rm 24\mu m})$ the Spitzer MIPS $24$~$\mu$m intensity 
in units of MJy\,sr$^{-1}$. $\dot{\Sigma}_{*}$ has the units of M$_{\odot}$kpc$^{-2}$yr$^{-1}$.
Following Helou et al. (2004), we subtracted the stellar continuum from the 24~$\mu$m surface brightnesses (in MJy\,sr$^{-1}$) using 
\begin{equation}
I_{\nu}(24\mu {\rm m})=I_{\nu}(24\mu {\rm m})-0.032\ I_{\nu}(3.6\mu {\rm m})\ .
\end{equation}
The full width at half-maximum (FWHM) of the point spread functions (PSFs), as stated in the Spitzer Observer's Manual 
(Spitzer Science Centre 2006), are 1.7 and 6~arcsec at 3.6 and 24~$\mu$m, respectively, that of the CO data is $10''$.
First, the data are convolved with kernels that match the PSFs of the images in the 3.6 and 24$\mu$m bands to a common PSF of $12''$.
The CO data of Gao et al. (2003) were also convolved to the common resolution.
Next, the data were re-binned to the common pixel size of $8.5''$.

\begin{figure}
  \centering
  \resizebox{\hsize}{!}{\includegraphics{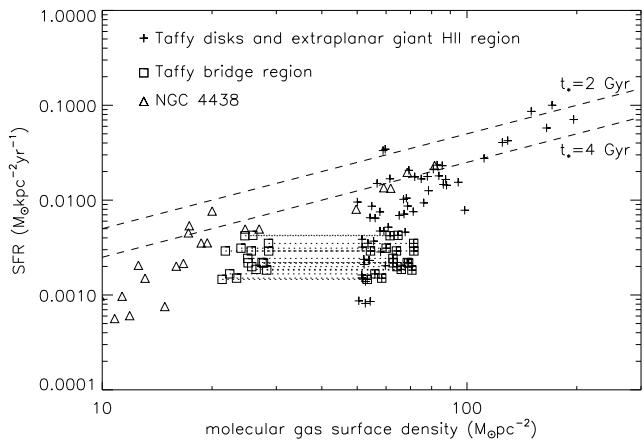}}
  \caption{Star formation rate based on the FUV and $24$~$\mu$m emission as a function of the molecular gas surface density.
   The crosses are based on $X=10^{20}$~cm$^{-2}$(K\,km\,s$^{-1}$)$^{-1}$, which is half of the Galactic value. 
   The bridge region outside the giant H{\sc ii}
   region near UGC~12915 are marked as squares with crosses. Squares without crosses correspond to a $2.5$ times lower $X$ factor as
   suggested by Zhu et al. (2007). Data of NGC~4438 are shown as triangles. The dashed lines correspond to constant
   star formation timescales of $2$ and $4$~Gyr.
  \label{fig:sfe}}
\end{figure}

The resulting pixel-by-pixel star formation rate as a function of the molecular gas surface density is presented in Fig.~\ref{fig:sfe}.
Since the CO-to-H$_{2}$ conversion factor $X$ is uncertain especially in the bridge region, we show molecular gas surface densities based on
$X=10^{20}$~cm$^{-2}$(K\,km\,s$^{-1}$)$^{-1}$ and $X=4 \times 10^{19}$~cm$^{-2}$(K\,km\,s$^{-1}$)$^{-1}$ (Zhu et al. 2007).
This should cover the expected range. The mean star formation rate timescale with respect to the molecular gas 
($t_{*} \sim  2$~Gyr; Bigiel et al. 2011) is shown as a dashed line. We also show a line with  $t_{*}=4$~Gyr, which is still in the expected range.
The star formation efficiency with respect to the molecular gas is normal in the galactic disks, but it is at least
a factor of $2$ to $3$ lower in the bridge region. This decrease in star formation efficiency is very similar to what is observed  
in the extraplanar gas of the Virgo spiral galaxy NGC~4438 (Vollmer et al. 2012b) and in the interacting galaxies NGC~3226/27
(region~2 in Table~1 of Lisenfeld et al. 2008). Whereas this decrease is observed at low molecular gas surface densities
($\sim 10$~M$_{\odot}$) in NGC~4438 and NGC~3226/27, it occurs at a $3$ to $5$ times higher molecular gas surface density in the Taffy bridge.
The quenching of star formation in the extraplanar gas of NGC~4438 is caused by a combined tidal and ram pressure interaction
(Vollmer et al. 2005).

The detection of CO(1--0) emission from the gas bridge implies high local gas densities of the order of 
$\ga 100$~cm$^{-3} = 5$~M$_{\odot}$pc$^{-3}$.
With an overall density of $0.01$~M$_{\odot}$pc$^{-3}$ on kpc-scales, this leads to a volume filling factor of $\la 2 \times 10^{-3}$.
The CO-emitting dense gas is thus highly clumped, as the disk ISM, and the clumps have a very low volume filling factor. 
We suggest that star formation is quenched by the high velocity dispersion of the bridge gas caused by mechanical energy input, 
which prevents the gravitational collapse of the high-density gas clumps. 

\begin{figure}
  \centering
  \resizebox{\hsize}{!}{\includegraphics{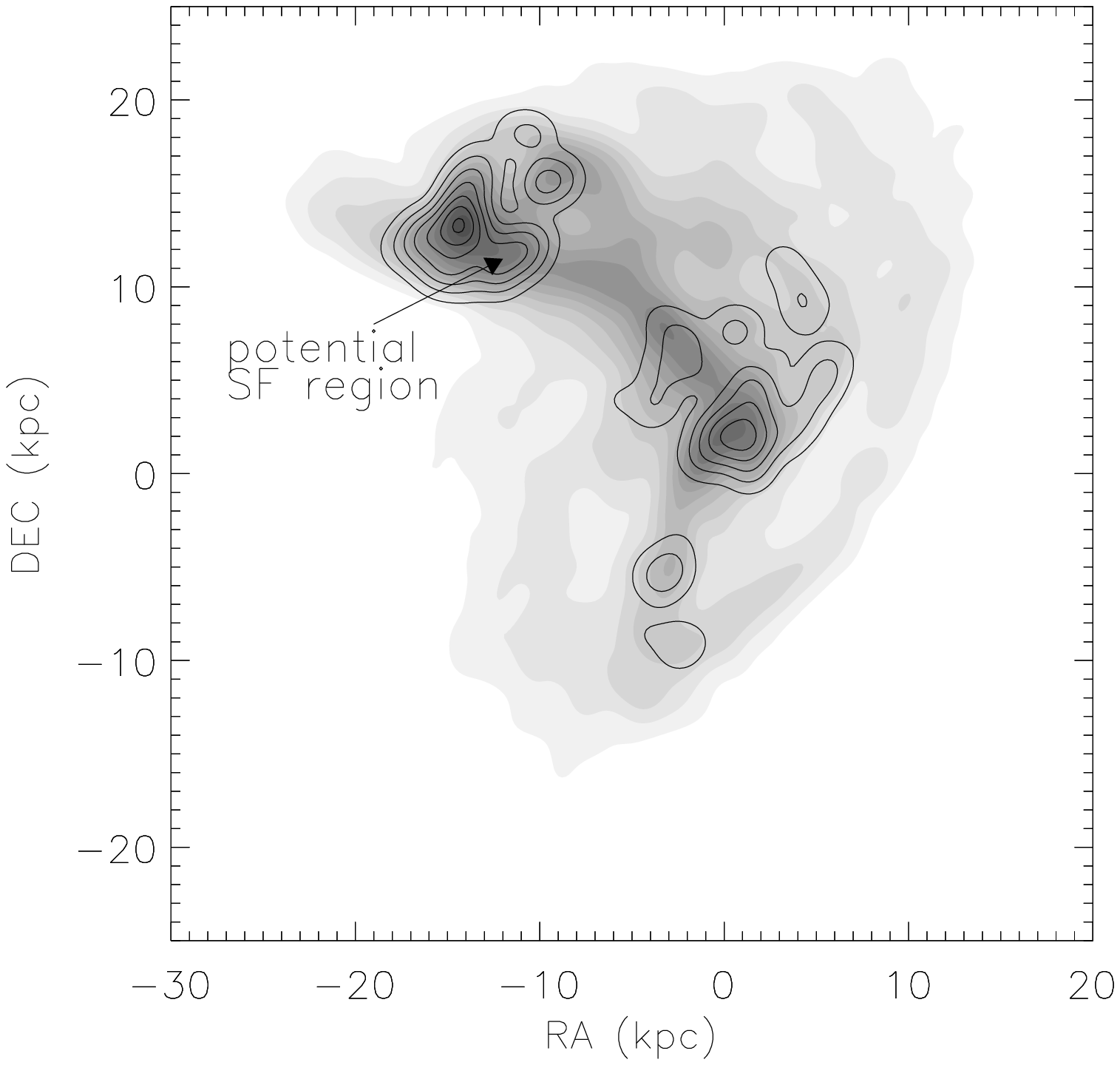}}
  \resizebox{\hsize}{!}{\includegraphics{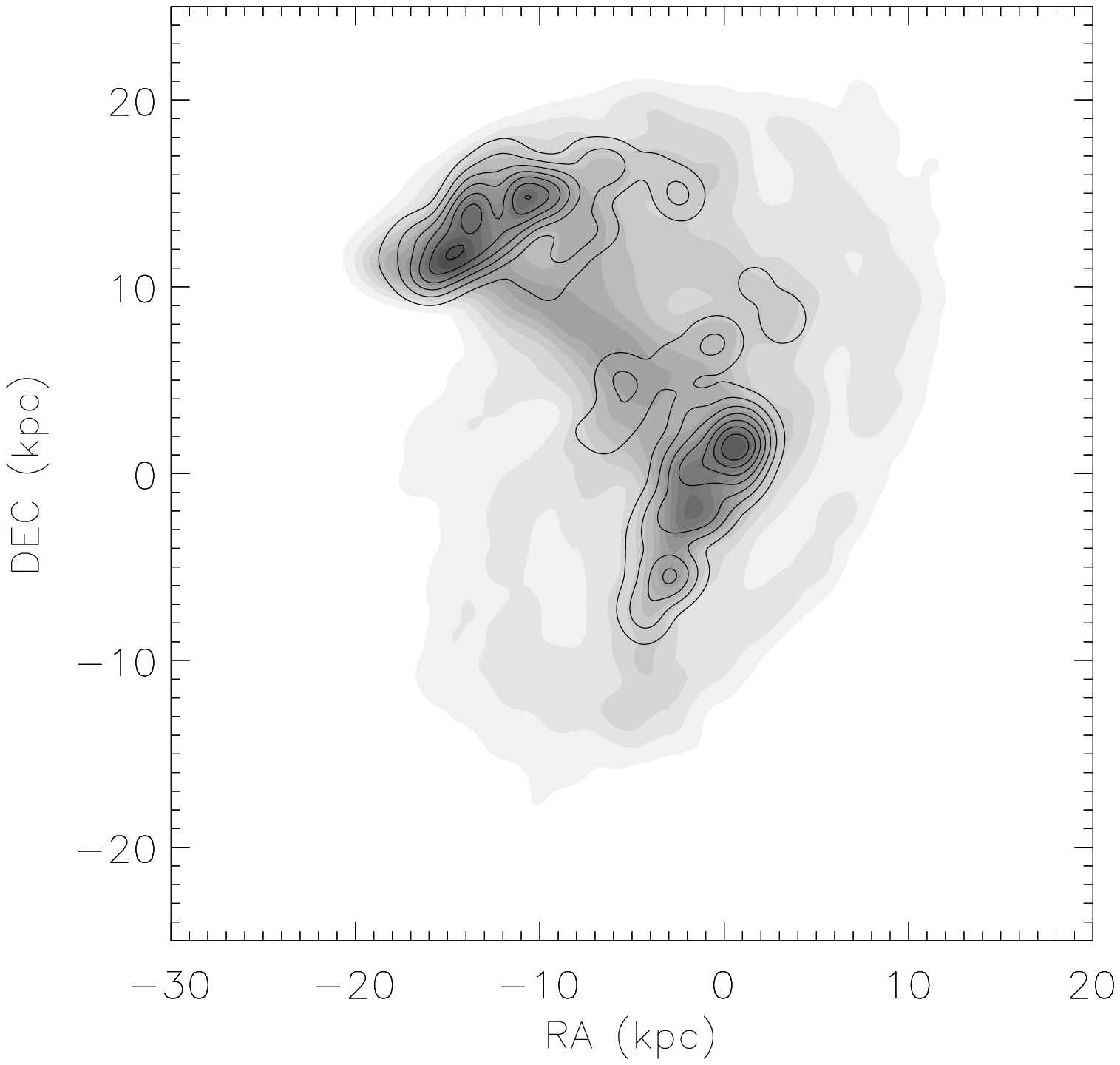}}
  \caption{Potential sites of star formation. Greyscale: total gas surface density. Contours: surface density of
    gas with a local Jeans length smaller than $2$~kpc. Upper panel: simulation 20. Lower panel: simulation 19.
  \label{fig:jeans}}
\end{figure}

Obviously, this suggestion does not hold for the giant H{\sc ii} region
near UGC~12915, where gas clouds do collapse and form stars. We suggest that this is caused by a higher local gas density
leading to a Jeans length which equals or is smaller than the extent of the gas.
To illustrate this effect, we calculated the local Jeans length for the gas in our simulations. In Fig.~\ref{fig:jeans} we
only show gas with a Jeans length smaller than $2$~kpc.
The gas in the bridges of both simulations has local Jeans lengths exceeding $2$~kpc. The gas north of the primary
galaxies' centers with local Jeans lengths smaller than $2$~kpc is mostly located in the disks of the primary galaxies.
Only in simulation 20 is there a region of high surface density south of the secondary galaxy's center at $(-12,12)$~kpc, which has 
a local Jeans length smaller than $2$~kpc. This region is supposed to form stars and might correspond to the
giant H{\sc ii} region south of UGC~12915. We do not claim to reproduce the giant H{\sc ii} region but we think
that Fig.~\ref{fig:jeans} can be seen as a proof of concept.

\section{Conclusions \label{sec:conclusions}}

The Taffy system UGC~12914/15 is a rare case of a head-on collision between two gas-rich spiral galaxies.
We made dynamical simulations of this galaxy collision, in which the gas of both galaxies collide, 
using a model which includes a collisionless (halo and stellar particles) and a collisional (gas) component. 
The gaseous component is simulated by a sticky particle scheme where gas particles undergo partially
inelastic collisions. The collision rate depends on the mass-radius relation of the gas particles (Eq.~\ref{eq:xi}).
With a first set of simulations with a simplified collision geometry, we investigated the behavior of the collision
rate as a function of cloud size parameter $\xi$. The collision rate increases with increasing $\xi$. 
Our collision rates are in reasonable agreement with theoretical expectation. A higher
collision rate leads to more momentum transfer between the cloud particles during the galaxy collision.
A higher collision rate during impact increases the gas mass of the bridge region between the two colliding galaxies.
We fixed $\xi$ to obtain a bridge gas mass exceeding $3 \times 10^{9}$~M$_{\odot}$.

In a second step, we made a second set of parameters with fixed inclination angles between the two counter-rotating
disks of $0^{\circ}$ and $-30^{\circ}$ where we varied systematically the position of impact.
We found two simulations, that reproduce the observed main characteristic of the Taffy system. For both simulations,
the relative velocity at impact is $\sim 1000$~km\,s$^{-1}$. The transverse velocity at the present time is between
$650$~km\,s$^{-1}$ and $660$~km\,s$^{-1}$. We confirm the time of impact estimated by Condon et al. (1993).
Our best-fit models place the impact $23$~Myr ago. For times $>20$~Myr all models develop ring structures characteristic
for galaxy head-on collisions.

To compare simulation snapshots to H{\sc i} and CO observations (Gao et al. 2003, Braine et al. 2003), 
we assume that the molecular fraction of the gas depends on the square root of the gas volume density. 
This recipe allows us produce model H{\sc i} and CO cubes. For the comparison of our simulations with observations
of polarized radio continuum emission, we calculated the evolution of the 3D large-scale magnetic field for our simulations.
The induction equations including the time-dependent gas-velocity fields from the dynamical model were solved for this purpose.

Since we could not find a single simulation which reproduces all observed characteristics, we present two ``best-fit'' simulations.
The first simulation better reproduces the H{\sc i} and CO line profiles of the bridge region (Braine et al. 2003), whereas the second simulation
better reproduces the stellar distribution of UGC~12915, the symmetric gas velocity fields of the galactic disks,
the projected magnetic field vectors in the bridge region, and the distribution of the 6~cm polarized radio continuum emission (Condon et al. 1993).
The following observational characteristics can be reproduced by our models:
\begin{enumerate}
\item
The stellar distribution of the Taffy system. The stellar distribution of the model secondary galaxy is more
distorted than that of UGC~12915.
\item
The prominent H{\sc i} and CO gas bridge.
\item
The offset of the CO emission to the south with respect to the H{\sc i} emission in the bridge region.
\item
The gas symmetric velocity fields in the galactic disks.
\item
The isovelocity contours of the CO velocity field which are parallel to the bridge.
\item
The H{\sc i} double-line profiles in the disk region.
\item
CO emission is only detected in the high-velocity component of the double-line. 
\item
The large gas linewidths ($100$-$200$~km\,s${-1}$) in the bridge region.
\item
The velocity separation between the double lines ($\sim 330$~km\,s$^{-1}$).
\item
The high field strength of the regular magnetic field in the bridge region.
\item
The projected magnetic field vectors, which are parallel to the bridge.
\item
The offset of the maximum of the 6~cm polarized radio continuum emission to the south of the bridge.
\item
The strong total power emission from the disk.
\end{enumerate}
The observed distortion of the H{\sc i} envelope of the Taffy system (Condon et al. 1993) cannot be reproduced by our model.
This is due to an insufficient momentum transfer between the gas particles located in the outer galactic disks.
For this gas a continuous description of the ISM (SPH or hydro) might be preferable.

The model allows us to redefine the bridge region in three dimensions. We estimate the total gas mass (H{\sc i}, warm and cold H$_{2}$)
to be $5$ to $6 \times 10^{9}$~M$_{\odot}$. Its molecular fraction $M_{\rm H_{2}}/M_{\rm HI}$ is about unity.
The structure of the model gas bridge is bimodal. There is a dense ($\sim 0.01$~M$_{\odot}$pc$^{-3}$) component
with a high velocity dispersion $> 100$~km\,s$^{-1}$ and a less dense ($\sim 10^{-3}$~M$_{\odot}$pc$^{-3}$)
component with a smaller, but still high velocity dispersion $\sim 50$~km\,s$^{-1}$. The synchrotron lifetime of
relativistic electrons is only long enough to be consistent with the existence of the radio continuum bridge (Condon et al. 1993) 
for the less dense component. This explains why the radio continuum emission follows the H{\sc i} and not the CO emission 
of the bridge outside the giant H{\sc ii} region near UGC~12915.
On the other hand, only the high-density gas undergoes a high enough mechanical energy input to produce the observed strong emission
of warm H$_{2}$ (Peterson et al. 2012). We propose that, despite the high local gas densities, 
this high input of mechanical energy quenches star formation in the bridge gas except for the giant H{\sc ii} region near UGC~12915.
Our model suggests that we observe this galaxy head-on collision near the time of maximum CO and H$_{2}$ emission.

\begin{acknowledgements}
We would like to thank Y.~Gao for providing his CO data to us.
This work has been supported by the Polish Ministry of Science and Higher Education grant No. 2011/03/B/ST9/01859.
\end{acknowledgements}

\appendix

\section{CO and H{\sc i} observed and model spectra of the Taffy system UGC~12914/15} 

For a better readability of this article, we decided to show the CO and H{\sc i} observed  and model spectra
as an appendix.

\begin{figure*}
  \centering
  \resizebox{\hsize}{!}{\includegraphics{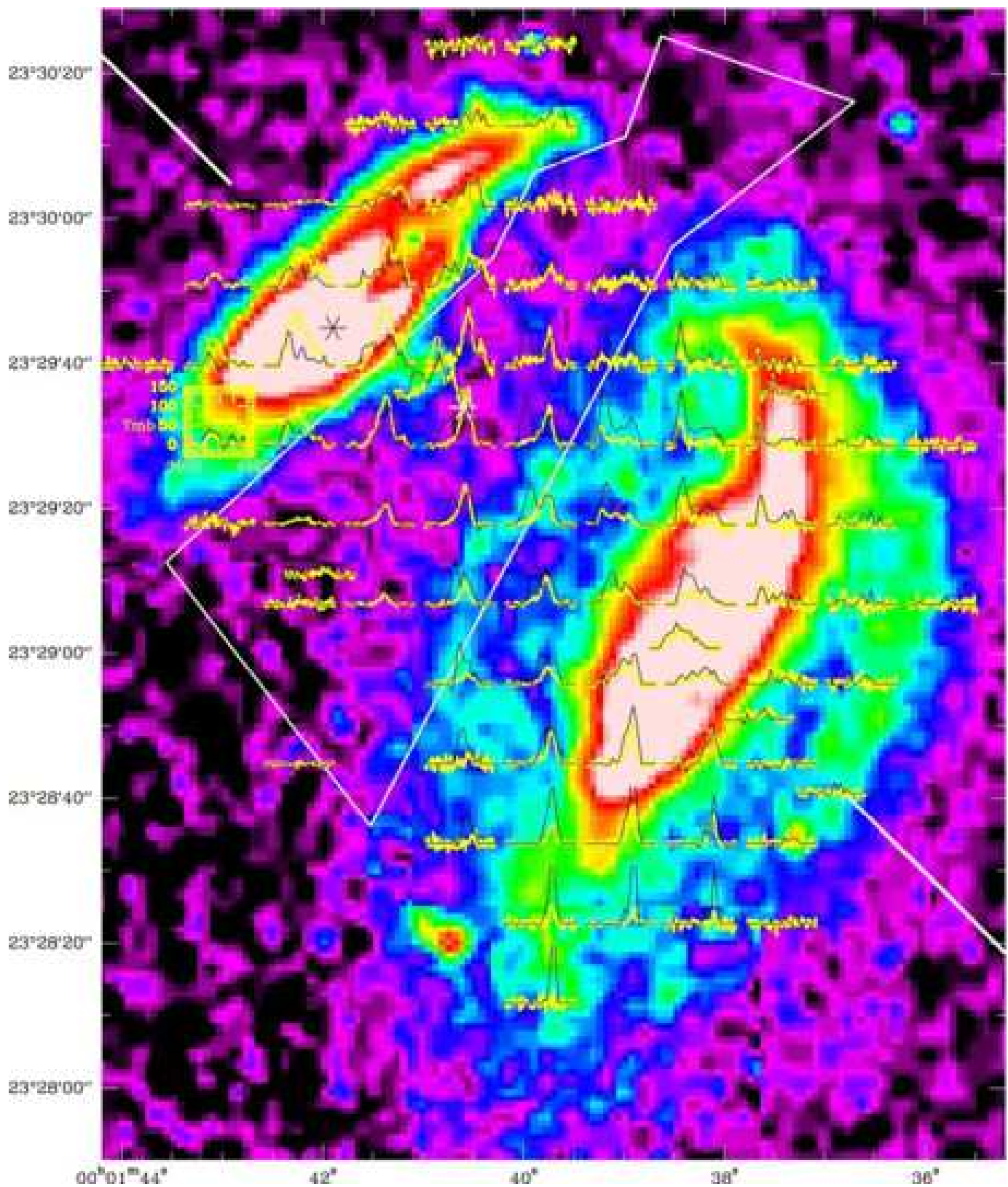}}
  \caption{CO(1-0) spectra (yellow line and yellow scale, intensity in milliKelvins) overlaid with H{\sc i} spectra 
    (black line) on a Digitized Sky survey image of the UGC 12914/5 system (from Braine et al. 2003). 
    The center of UGC 12915 is at $00^{\rm h}01^{\rm m}41.9^{\rm s}$, $23^\circ 29'44.9''$. 
    The bridge region is delimited by the white polygon.
  \label{fig:img20}}
\end{figure*}

\begin{figure*}
  \centering
  \resizebox{\hsize}{!}{\includegraphics{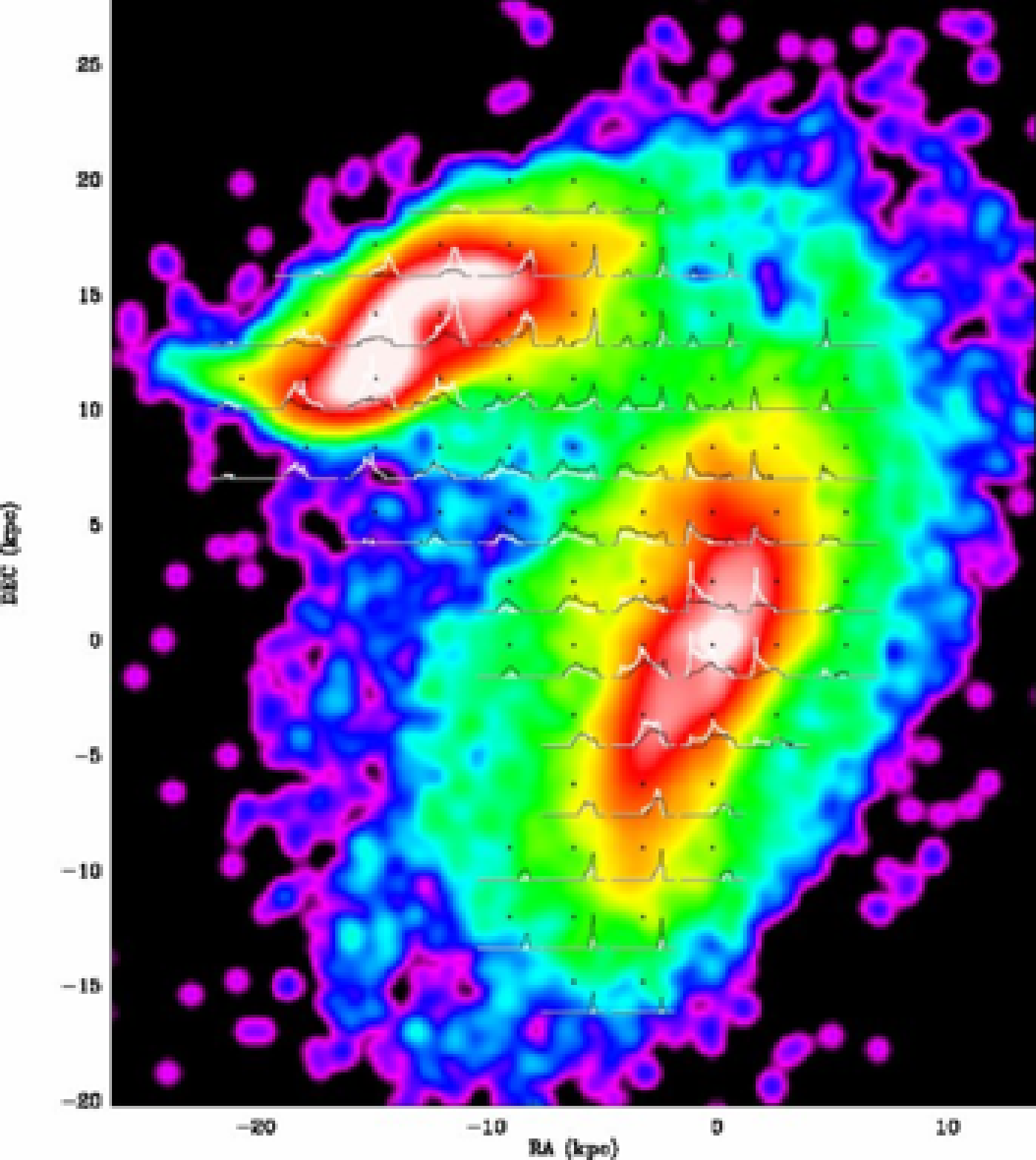}}
  \caption{Simulation 20. CO(1-0) model spectra (white line) overlaid with H{\sc i} spectra 
    (black line) on the model stellar surface density distribution.
  \label{fig:taffy26new2_az20}}
\end{figure*}
\begin{figure*}
  \centering
  \resizebox{\hsize}{!}{\includegraphics{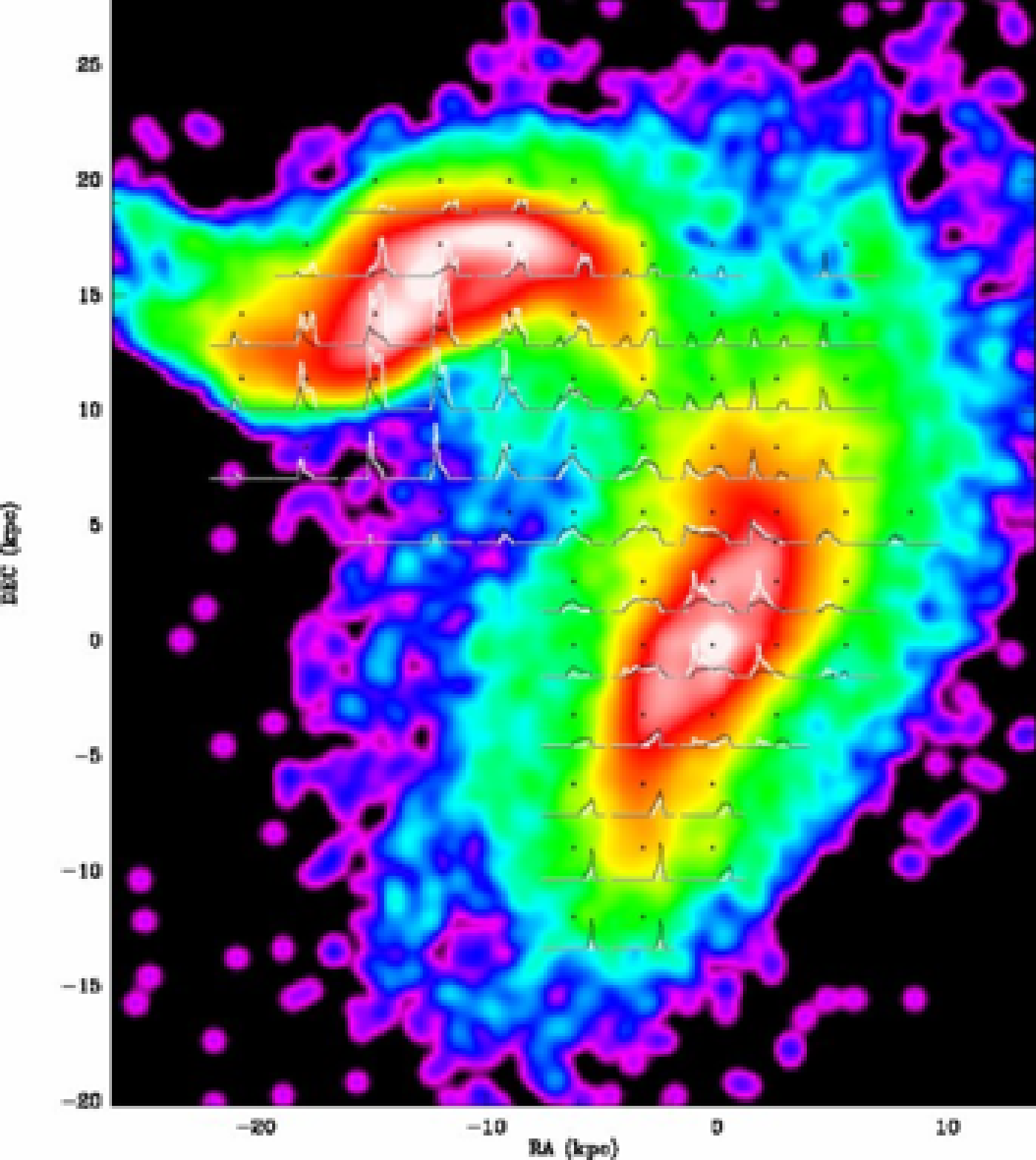}}
  \caption{Simulation 19. CO(1-0) model spectra (white line) overlaid with H{\sc i} spectra 
    (black line) on the model stellar surface density distribution.
  \label{fig:taffy22new_az20}}
\end{figure*}

\begin{figure*}
  \centering
  \resizebox{\hsize}{!}{\includegraphics{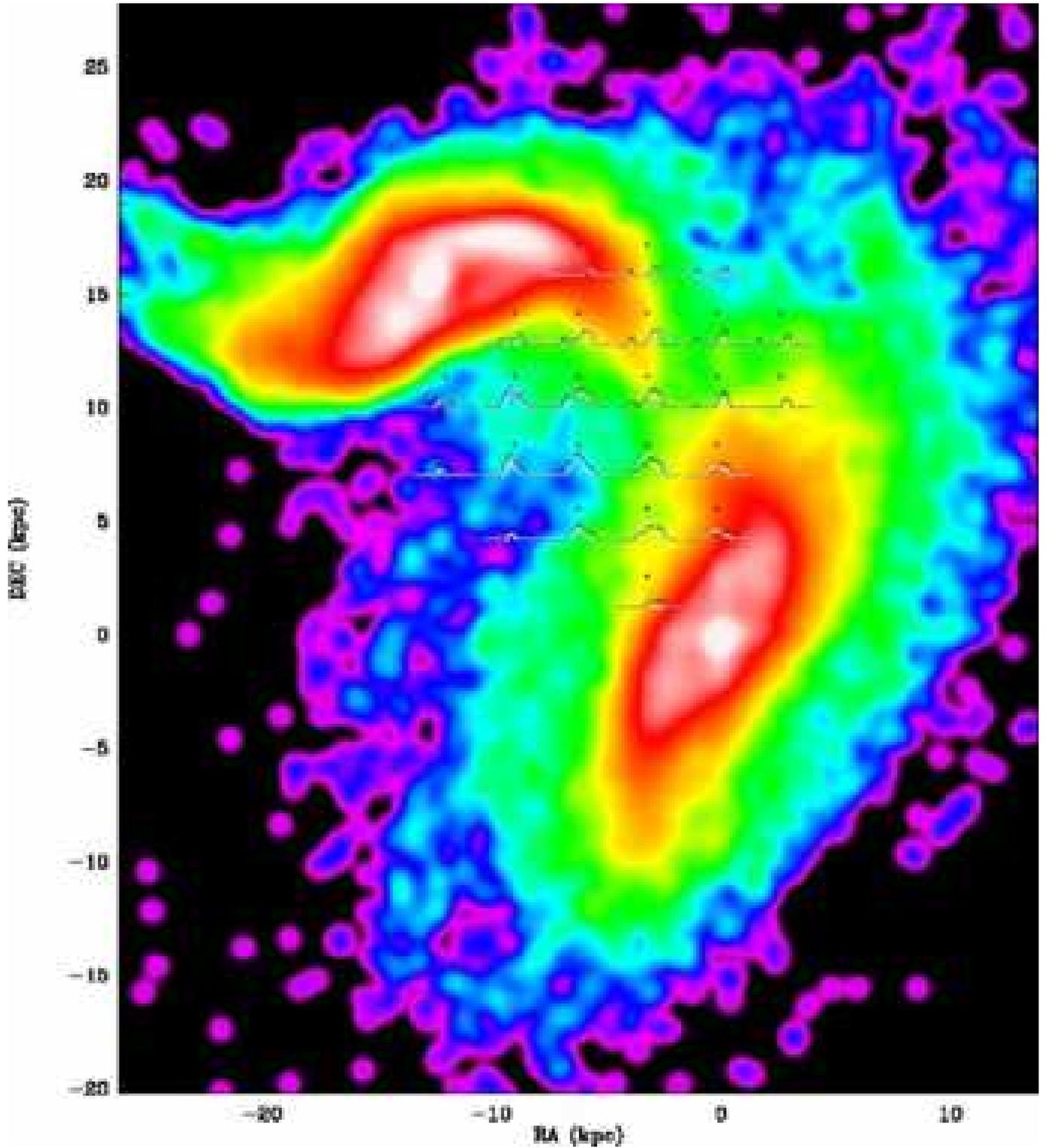}}
  \caption{Simulation 19 where the disk gas has been removed. Thus only the gas located in the
    bridge region is visible. CO(1-0) model spectra (white line) overlaid with H{\sc i} spectra 
    (black line) on the model stellar surface density distribution.
  \label{fig:taffy22new_az20_bridge}}
\end{figure*}

\begin{figure*}
  \centering
  \resizebox{\hsize}{!}{\includegraphics{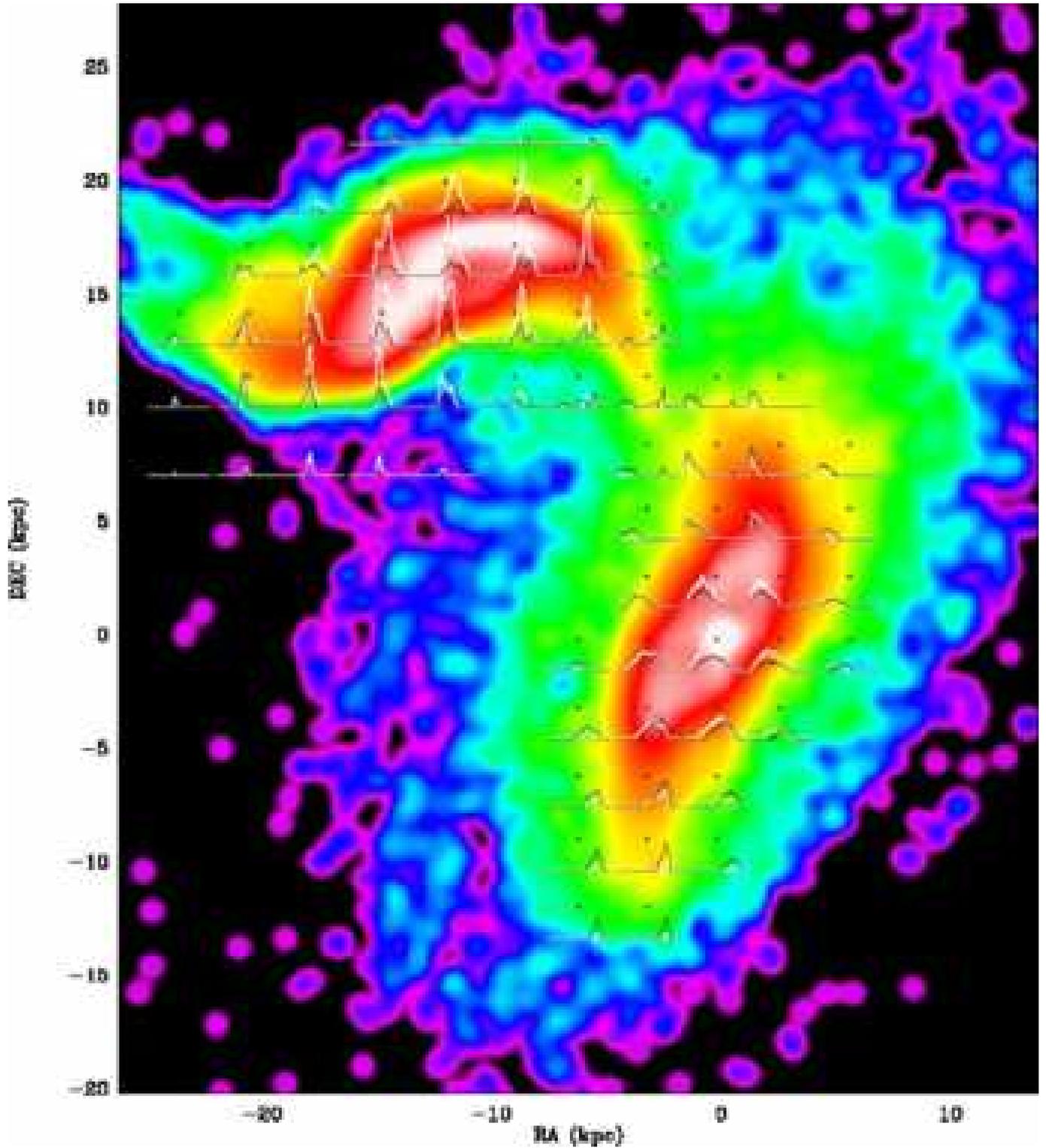}}
  \caption{Simulation 19 wihout cloud-cloud collisions. CO(1-0) model spectra (white line) overlaid with H{\sc i} spectra 
    (black line) on the model stellar surface density distribution.
  \label{fig:taffy22new_nocoll_az20}}
\end{figure*}

\end{document}